\documentclass[11pt]{article}

\usepackage[ruled, vlined, linesnumbered, noresetcount]{algorithm2e}
\usepackage{times}
\usepackage{amsthm}
\usepackage{amsmath}
\usepackage{amsfonts}
\usepackage{amssymb}
\usepackage{enumerate}
\usepackage{graphicx}
\usepackage{color}
\usepackage{multicol}
\usepackage{color}
\usepackage{authblk}

\DeclareMathOperator*{\argmax}{arg\,max}

\SetKwFor{ForAll}{forall}{do}{end forall}

\newtheorem{theorem}{Theorem}
\newtheorem{lemma}{Lemma}

\newtheorem{prop}{Proposition}

\newtheorem{example}{Example}

\def\proof{\noindent{\bf Proof. $\ $}}
\def\qed{\hfill $\Box$}

\def\s{{\sigma}}
\def\d{{\delta}}
\def\bs{{\bf s}}
\def\cs{s}
\def\Wbs{C(\bs)}
\def\Wss{C(\bs^*)}

\newcommand{\remove}[1]{}

\def\T{{{\cal T}}}
\def\H{{{\cal K}}}
\def\G{{{\cal G}}}
\def\vj{{v_j}}

\def\S{{S}}

\def\c{{ C}}
\def\gad{{\Lambda}}
\newcommand\myatop[2]{\genfrac{}{}{0pt}{}{#1}{#2}}

\def\D{{\rm  B}}
\def\X{{ X}}

\newcommand{\N}{{\mathbb{N}}}
\newcommand{\Active}{{{\sf Active}}}

\begin{document}

\title{Optimizing Spread of Influence in Weighted  Social Networks via Partial Incentives\thanks{
A preliminary version of this paper was presented at the 22nd International Colloquium on Structural Information and Communication Complexity (SIROCCO 2015), Montserrat, Spain, July 15 - 17, 2015}}

\author[1]{Gennaro Cordasco}
\author[2]{Luisa Gargano}
\author[2]{Adele~A.~Rescigno}
\author[2]{Ugo Vaccaro}

\affil[1]{Department of Psychology, Second University of Naples, Italy, \\{\tt {gennaro.cordasco@unina2.it}}}
\affil[2]{Department of Computer Science, University of Salerno, Italy, \\{\tt{\{lgargano,arescigno,uvaccaro\}@unisa.it}}}

\maketitle
\begin{abstract}
A widely studied process of influence diffusion in social networks
posits that the dynamics  of influence diffusion 
evolves as follows: Given a graph  $G=(V,E)$, representing the network, 
initially \emph{only} the members of a given  $S\subseteq V$   are influenced; 
subsequently, at each round, the set of influenced nodes is augmented 
by all the nodes in the network  that have a sufficiently large number 
of already influenced neighbors. The general problem is to find a small initial 
set of nodes that influences the whole network.
In this paper we extend  the previously described  basic model in the following ways:
firstly,   we assume that 
there are non negative values $c(v)$  associated to 
each node $v\in V$, measuring   how much
it costs to initially influence node $v$,  and the algorithmic problem is
to find a set of nodes of \emph{minimum total cost} that influences the whole network;
successively, we study the consequences of giving 
\emph{incentives} to member of the networks, and we quantify how 
this affects (i.e., reduces) the total costs of starting 
process that influences the whole network. For the two  above 
problems we provide  both hardness  and algorithmic results.
We also experimentally  validate our algorithms via extensive simulations on real 
life networks.

\noindent\textbf{Keywords:} Social Networks; Spread of Influence; Viral Marketing 
\end{abstract}


\section{Introduction}
Social influence is the process by which individuals adjust their opinions, 
revise their beliefs, or change their behaviors as a
result of  interactions with other people.
It has not escaped the attention of advertisers
that the natural  human tendency to conform
can be exploited in \emph{viral marketing} \cite{LAM}.
Viral marketing 
refers to the spread of information about products and  behaviors,
  and their adoption by people.
  For what strictly concerns us,  the intent of maximizing the spread of viral information across a network naturally
suggests many interesting  optimization  problems. Some of them  were first articulated in the seminal papers
\cite{KKT-03,KKT-05}, under various adoption paradigms.
The recent monograph \cite{CLC} contains an excellent  
description of the area.
In the next section, we will explain and motivate our model of 
information diffusion, state the problems that we plan to investigate,
describe our results, and discuss how
they relate to the existing literature.

\subsection{The Model}
Let $G = (V,E)$ be a graph modeling a social network. 
We denote by $\Gamma_G(v)=\{u\in V : (v,u)\in E\}$ and by $d_G(v)=|\Gamma_G(v)|,$ 
respectively, the neighborhood and the degree of vertex $v$ in $G$. 
Let $S \subseteq V$, and let $t: V \to\N = \{1,2,\ldots\}$ be a
function assigning integer thresholds to the vertices of $G$;
we  assume w.l.o.g. that $1\le t(u)\le d(u)$ holds for all  $v\in V$. 
For each node $v\in V$, the value 
 $t(v)$  quantifies  how hard it is to influence  node $v$, in the sense that
 easy-to-influence  elements of the network  have ``low'' $t(\cdot)$ values, and
hard-to-influence  elements have  ``high'' $t(\cdot)$ values \cite{Gr}.
An {\em activation process in $G$ starting at $S\subseteq V$}
is a sequence  

$$\Active_G[S,0] \subseteq \Active_G[S,1] 
\subseteq \ldots\subseteq \Active_G[S,\ell] \subseteq \ldots \subseteq V$$

\noindent
of vertex subsets\footnote{We will omit the 
subscript $G$ whenever the graph $G$ is clear from the context.}, with
$\Active_G[S,0] = S$, and such that for all $\ell > 0$,

$$\Active_G[S,\ell] = \Active_G[S,\ell-1]\cup \Big\{u \,:\, \big|\Gamma_G(u)\cap \Active_G[S,\ell - 1]\big|\ge t(u)\Big\}.$$

\noindent
In words, at each round  $\ell$ the set of active (i.e, influenced) nodes  is augmented by the set of  nodes  $u$ that have a number of
\emph{already} activated   neighbors greater or equal to  $u$'s threshold $t(u)$. 
We say that $v$ {\em is activated} at round $\ell>0$ if $v \in  \Active_G[S,\ell]- \Active_G[S,\ell - 1]$.
A target set for $G$ is a set $S$  that  will activate the whole network, that is,  $\Active_G[S,\ell]=V$, for some $\ell \geq 0$.
The classical  Target Set Selection (TSS) problem (see e.g. \cite{ABW-10,Cic+})
  is defined as follows:

\medskip
 {\sc Target Set Selection}.

{\bf Instance:} A network $G=(V,E)$ with  thresholds $t:V\longrightarrow \mathbb{N}$.
 
{\bf Problem:} Find a  target set $S\subseteq V$ of \emph{minimum} size for $G$.

\medskip
\noindent
The TSS Problem has roots in the general study
of the \emph{spread of influence} in Social Networks (see  \cite{CF,CLC,EK}).
For instance, in the area of viral marketing \cite{DR-01}, companies  wanting to
promote products or behaviors might  initially  try to target and convince
a set of individuals (by offering  free copies of the products or some 
equivalent monetary rewards) who, by word-of-mouth, can  successively trigger
a  cascade of influence in the network leading to
an  adoption  of the products by  a much larger number of individuals.
In order to make the model more realistic, 
we extend the previously described  basic model in two ways: First, we assume that 
there are non negative values $c(v)$  associated to 
 each vertex $v\in V$, measuring   how much
it costs to initially convince the member $v$ of the network
to endorse a given product/behavior. Indeed, 
that different members of the network have different
activation costs (see \cite{Ba+}, for example) is  justified by
the observation  that celebrities or public figures can  charge more for their
endorsements of  products. Therefore, we are lead to our first extension of 
the TSS problem:

\medskip

{\sc  Weighted Target Set Selection  (WTSS)}.

{\bf Instance:} A network $G=(V,E)$, thresholds $t:V\to \mathbb{N}$,  costs
$c:V\to \mathbb{N}$.

{\bf Problem:} Find a  target set $S\subseteq V$ of \emph{minimum} cost  $\c(S)=\sum_{v\in S}c(v)$ 

 among all target sets for  $G$.

\medskip
\noindent
Our second, and more technically challenging, extension of the classical TSS problem is inspired by
the  recent interesting paper \cite{Dem14}. 
In that paper the authors observed that the basic model misses a
crucial feature of practical applications since it forces
the optimizer to make a binary choice of either zero or 
complete influence on  each individual (for example, either not offering or 
offering a free copy of the product to  individuals in order to  initially
convince them to adopt the product and influence their friends about it). 
In realistic
scenarios, there could be more reasonable and effective options. 
For example, a company
promoting a new product may find that offering for free ten copies
of a product is far less effective than offering a discount
of ten percent to a hundred of people. Therefore, we formulate 
our second extension of the basic model as follows.

\remove{
\begin{example}\label{example1}
Consider the tree $T$ in figure \ref{fig:WTSS}. Both
$S=\{v_2,v_3, v_4,v_5, v_6\}$ and $S'=\{v_2,  v_9, v_{10},v_{12},v_{13},v_{15},v_{16},v_{18},v_{19}\}$ are target sets  for    $T$. 
Indeed we have
{\begin{eqnarray*}
 \Active[S,0] &=& \{v_2,v_3, v_4,v_5, v_6\}, \\
	\Active[S,1]&=& \Active[S,0]\cup\{v_1, v_7, v_8, \ldots,v_{20}\}=V\\ 
		\\
\Active[S',0] &=&  S', \\
\Active[S',1]&=&  \Active[S',0]\cup\{v_1,v_4,v_5,v_6,v_7,v_8\}, \\
 \Active[S',2]&=&  \Active[S',1]\cup \{ v_3,  v_{14}, v_{17} , v_{20}  \},   \\
\Active[S',3]&=&  \Active[S',2]\cup \{  v_{11}  \}   =V.
\end{eqnarray*}
}

\noindent
We notice that, while $S$ is a minimum size target set for the graph $T$ and 
 $|S|=5<9=|S'|$,  the cost of $S$ is greater than the cost of $S'$,  $\c(S)= 18>14=\c(S')$.
\end{example}
\begin{figure}
\begin{center}
\includegraphics[height=3truecm]{albero-inc.eps} 
\caption{A tree example. The number inside each circle denotes the vertex threshold.\label{fig:WTSS}}
\end{center}
\end{figure}
}

\medskip
\noindent
{\bf Targeting with Partial Incentives.}
An assignment of  partial incentives  to the vertices 
of a network $G=(V,E)$, with  $V=\{v_1, \ldots , v_n\}$, is a vector  
$\bs=(\cs(v_1), \dots , \cs(v_n))$,  where 
$\cs(v)\in  \N_0 = \{0,1,2,\ldots\}$ represents the amount of influence we initially apply on 
$v\in V$. The effect of applying incentive $\cs(v)$ on node $v$ is to decrease its 
 threshold, i.e., to make individual $v$ more susceptible to future influence.
 It is clear that to start the process, there should be a sufficient number of  nodes $v$'s to
 which the amount of exercised influence $\cs(v)$ is at least equal to their 
 thresholds $t(v)$. Therefore, 
an {\em activation process in $G$ starting with incentives whose values are given by the vector $\bs$}
is a sequence  of vertex subsets
		$$\Active[\bs,0] \subseteq \Active[\bs,1] \subseteq \ldots
                               \subseteq \Active[\bs,\ell] \subseteq \ldots \subseteq V,$$
with
$\Active[\bs,0] = \{v \ | \ \cs(v)\geq t(v)\}$, and such that for all $\ell>0$,
			$$\Active[\bs,\ell] = 
			 \Active[\bs,\ell-1]\cup \Big\{u \,:\, \big|\Gamma_G(u)\cap \Active[\bs,\ell - 1]\big|\geq t(u)-s(u)\Big\}.$$
A \emph{target vector} $\bs$  is an  assignment of partial incentives 
that triggers an activation process influencing the whole network, that is, 
such that $\Active[\bs,\ell]=V$ for some  
$\ell \geq 0$.
The  Targeting with  Partial Incentive problem  can be  defined as follows:

\medskip

 {\sc Targeting with Partial Incentives (TPI)}.

{\bf Instance:} A network $G=(V,E)$, thresholds $t:V\longrightarrow \mathbb{N}$.

{\bf Problem:} Find target vector $\bs$  which minimizes  $\Wbs=\sum_{v\in V} \cs(v)$.

\medskip
\noindent
Notice that the  Weighted Target Set Selection problem, when the costs $c(v)$ are always
equal to the thresholds $t(v)$, for each $v\in V$, can be seen as 
a particular case of Targeting with Partial Incentives in which 
the incentives $\cs(v)$ are set either to $0$ or to $t(v)$. 
Therefore, in a certain sense, the Targeting with Partial Incentives
can be seen as 
 a kind of ``fractional" counterpart of the Weighted Target Set Selection  problem
(notice, however, that the incentives $\cs(v)$  are integer as well).
In general,  the two optimization problems are quite
different since  arbitrarily large gaps are possible between the costs of 
 the solutions of the WTSS and TPI problems, as the following example shows.

\begin{example}\label{example2}
Consider the complete graph on $n$ vertices $v_1,\ldots,v_n$, with thresholds 
$t(v_1)=\ldots= t(v_{n-2})=1$,   $t(v_{n-1})=t(v_{n})=n-1$ and costs 
equal to the thresholds.
An optimal  solution to the WTSS problem   consists of either 
vertex $v_{n-1}$ or vertex  $v_{n}$, 
 hence  of total cost equal to $n-1.$ On the other hand, if 
 partial incentives are possible one can  assign incentives
$\cs(v_1)=\cs(v_{n})=1$ and $\cs(v_i)=0$ for  $i=2,\ldots,n-1$, and have an optimal solution of 
value equal to 2.
Indeed, we have 

\begin{itemize}
\item $\Active[\bs,0]=\{v_1\}$, since $t(v_1)=\cs(v_1)$,
\item $\Active[\bs,1]=\{v_1,v_2,\ldots,v_{n-2}\}$, since $t(v_i)=1$ for $i=2,\ldots,n-2$,
\item $\Active[\bs,2]=\{v_1,v_2,\ldots,v_{n-2},v_n\}$, since $t(v_n)-\cs(v_n)=n-2$, and 
\item $\Active[\bs,3]=\{v_1,v_2,\ldots,v_{n-1},v_n\}$, since $t(v_{n-1})=n-1$.
\end{itemize}

\noindent
Hence, an optimal  solution to the WTSS problem $S^*$  has  $\c(S^*)=n-1$
while an optimal  vector $\bs^*$ has  
$\Wss=\sum_{i=1}^n \cs^*(v_i)= 2$ 
independent of $n$.
\end{example}

\subsection{Related Works} 
The  algorithmic problems we have articulated have roots in the general study
of the \emph{spread of influence} in Social Networks (see  \cite{CLC,EK} and references quoted therein).
 The first authors to  study problems of spread of influence in networks from an algorithmic point of view were Kempe \emph{et al.} \cite{KKT-03,KKT-05}.{ They introduced the Influence Maximization problem, where the goal is to identify a set $S\subseteq V$ such that its cardinality is bounded by a certain budget $\beta$ and the activation process activates as much vertices as possible.}
However, they were mostly interested in networks with  randomly chosen thresholds.
Chen \cite{Chen-09} studied the following minimization problem:
Given a graph $G$ and fixed arbitrary thresholds $t(v)$, $\forall v\in V$, find
a target set of minimum size that eventually activates
all (or a fixed fraction of) nodes of $G$.
He proved  a strong   inapproximability result that makes unlikely the existence
of an  algorithm with  approximation factor better than  $O(2^{\log^{1-\epsilon }|V|})$.
Chen's result stimulated a series of papers  
(see for instance \cite{ABW-10,BCNS,BHLM-11,Centeno12,Chang,Chun,Chun2,Chopin-12,Cic+,Cic14,C-OFKR,Fr+,Ga+,Gu+,Li+,Mo+,NNUW,Re,W+,Za} and references therein quoted)
that isolated 
many interesting scenarios 
in which the problem (and variants thereof) become tractable.
{ The Influence Maximization problem 
with partial incentives was introduced in \cite{Dem14}. In this model the authors assume that the thresholds are randomly chosen values in 
the interval $(0,1)$ and they aim to understand how a fractional version of the Influence Maximization problem differs from the original version. 
{To that purpose}, they introduced the concept of partial influence and showed that, from a theoretical point of view, the fractional version retains essentially the same computational hardness as the integral version. However,  from  the practical side,  the  authors of \cite{Dem14}
proved that it is possible to efficiently compute   solutions in the fractional setting,
whose costs are much smaller than 
the best solutions to  the integral version of the problem.
We point out that 
the model in \cite{Dem14} assumes the existence  of  functions $f_v(A)$ that 
quantify the influence of arbitrary subsets of vertices $A$ on each vertex $v$.
 In the widely studied  ``linear threshold''  model,  a vertex is  influenced by its neighbors \emph{only}, and such neighbors have the same
influencing power  $f_v$ on $v$; this is  equivalent to the model considered in this paper.
Indeed, a $WTSS$ instance with threshold function $t:V\rightarrow \mathbb{N}$ can be  transformed into an instance with threshold  function $t':V\rightarrow (0,1)$ by setting $t_{max}>\max_{v\in V} t(v)$,  $t'(v)=t(v)/t_{max}$,  
  and  $f_v=1/t_{max}$, for each vertex $v\in V$.
	The viceversa  holds by setting $t(v)=\lceil t'(v)/f_v\rceil$, for each $v\in V$. 

\subsection{Our Results}
Our main contributions are the following. We first show, 
in Section \ref{hard},
that there exists a (gap-preserving) reduction from the classical
TSS problem to our  TPI and  WTSS problems (for the WTSS problem, the 
gap preserving reduction holds
also in the  particular case in which 
$c(v)=t(v)$, for each $v\in V$).  Using the important 
results by \cite{Chen-09}, this implies 
the TPI and  WTSS problems cannot be approximated to within a ratio 
of  $O(2^{\log^{1-\epsilon} n})$,  for any fixed $\epsilon>0$, 
unless $NP\subseteq DTIME(n^{polylog(n)})$ (again, for the latter problem this 
inapproximability result holds also in the case $c(v)=t(v)$, for each $v\in V$). 
Moreover, since the WTSS problem  is equivalent to the TSS problem when all thresholds are equal,
the reduction also show that the particular case in which
$c(v)=t(v)$, for each $v\in V$,  of the WTSS problem  is NP-hard. Again,
this is due to the corresponding hardness result of TSS given in \cite{Chen-09}.

{ In Section \ref{sec:WTSS} we present
a polynomial time algorithm {that}, given a  network and vertices thresholds,  { computes} 
 a cost efficient target set.} Our  polynomial time algorithm exhibits the following features: 
\begin{enumerate}
	\item For general graphs, 
it always returns a solution of cost at most equal to $\sum_{v\in V} \frac{  c(v)t(v)}{d_G(v) +1}$.
It is interesting to note that, when $c(v)=1$ for each $v\in V$, we recover 
the same upper bound on the cardinality of an optimal target set
given  in  \cite{ABW-10}, and proved therein  by means of the probabilistic method.
\item For  complete graphs our algorithm always returns a solution of \emph{minimum} cost.
\end{enumerate}

 In Section \ref{sec:TPI} we turn our attention to the  problem of spreading of influence with incentives  and we propose  a 
 polynomial time algorithm that, given a  network and vertices thresholds,   computes  a cost efficient target vector.  Our  algorithm exhibits  the following 
features: 
\begin{enumerate} 
\item For general graphs, 
it always return a solution $\bs$ (i.e., a target vector)   for $G$ of cost
$\Wbs=\sum_{v\in V}\cs(v)\leq \sum_{v\in V}  
\frac{t(v)(t(v) +1)}{2(d_G(v)+1)}$.
\item For trees and complete graphs our algorithm always returns an  \emph{optimal} target vector.
\end{enumerate}

Finally, in Section \ref{sec:experiments} we experimentally validate our algorithms
by running them on real life networks, and we compare the obtained results
with that of well known heuristics in the area (especially tuned  to our scenarios).
The experiments  shows that our algorithms consistently outperform those heuristics.

\section{Hardness of WTSS and TPI}\label{hard}  
We shall prove the following result.
\begin{theorem}
WTSS and TPI cannot be approximated within a ratio 
of  $O(2^{\log^{1-\epsilon} n})$ for any fixed $\epsilon>0$, 
unless $NP\subseteq DTIME(n^{polylog(n)})$.
\end{theorem}
\proof
We first construct  a gap-preserving reduction from the  TSS problem. 
The claim of the theorem follows from the inapproximability of TSS 
proved in \cite{Chen-09}. In the following, we give the full technical details    only for the    TPI problem.

Starting from an arbitrary  graph  $G=(V,E)$ with threshold function $t$,
input instance  of 
the TSS problem, we build a  graph $G'=(V', E')$ as follows:

\medskip
\noindent
\begin{itemize}
\item $V'=\bigcup_{v\in V}V'_v$ where $V'_v=\{v',v'',v_1,\ldots v_{d_G(v)}\}$. 
		 In particular, 
\begin{itemize}
		\item  we replace each  $v\in V$ by the gadget $\gad_v$ (cfr. Fig. \ref{fig:gadget}) in which the vertex set is  $V'_v$ and  $v'$ and  $v''$  are connected by the 
		disjoint paths  ($v',v_i,v''$) for $i=1,\ldots, d_G(v)$; 
	\item the threshold of $v'$ in $G'$ is equal to the threshold  
	$t(v)$ of $v$ in $G$, while each other vertex in $V'_v$ has threshold equal  to 1.
	\end{itemize}	
\noindent
$\bullet$    $E'=\{ (v',u')\ |\  (v,u)\in E\} \cup \bigcup_{v\in V} \{(v',v_i) , (v_i,v'')\mbox{, for  } i=1,\ldots, d_G(v)\}$.
\end{itemize}
\medskip
Summarizing,  $G'$ is constructed in such a way that for each gadget $\gad_v$, the vertex $v'$ plays the role of $v$ and is connected to all the gadgets representing neighbors of $v$ in $G.$ 
Hence, $G$ corresponds to  the  subgraph of $G'$ induced by the set $\{v'\in V'_v |\  v \in V\}.$
It is worth  mentioning that during an activation process if any vertex that belongs to a gadget  $\gad_v$ is active, 
then all the vertices in $\gad_v$ will be activate within the next  $3$ rounds.
\\
We claim that there is a target set $S \subseteq V$ for $G$ of cardinality $|S|=k$ \emph{if and only} if there is a target vector $\bs$ for $G'$ and $\Wbs= \sum_{u\in V'} \cs(u)=k$.
\\
Assume that $S\subseteq V$ is a target set for $G$, we can easily build an assignation  of partial incentives $\bs$ as follows:
$$
\centerline{$s(u)= \begin{cases} 1 & \mbox{ if  $u$ is the extremal vertex $v''$ in the gadget $ \gad_v$  and $v \in S$;}   \\ 
0 & \mbox{ otherwise.} \end{cases}$}
$$
Clearly,  $\Wbs=\sum_{v\in S} 1 =|S|$. To see that $\bs$ is a target vector we notice that\\
$\Active_{G'}[\bs,2]= \{u \ | \ u \in V'_v, v\in S\}$, consequently since $S$ is a target set and  $G$ is 
isomorphic to the  subgraph of $G'$ induced by  $\{v'\in V'_v |\  v \in V\}$, all the vertices  $v \in V'$ will be activated.
\\
On the other hand, assume that $\bs$ is a target vector for $G'$  and $\Wbs=k$, we can easily build a target set $S$ 
$$
\centerline{$S=\{v \in V \ | \  \exists u \in V'_v \mbox{ such that } s(u)>0 \}$.  } \vspace{-0.2truecm}
$$
By construction $|S| \leq \sum_{u\in V'} s(u)= \Wbs$. To see that $S$ is a target set for $G$, for each $v\in V$ we consider two cases on the values $\cs(\cdot)$:
\\  If there exists $u\in V'_v$ such that $\cs(u)>0$ then, by construction $v \in S$.
\\
 Suppose otherwise  $\cs(u)=0$ for each  $u\in V'_v$. We have that   in order to activate $v'$ (and then 
	any other vertex in  $\gad_v$) there must exist a round $i$ such that
	$\Active_{G'}[\bs,i-1]\cap (V'-V'_v)$ contains $t(v)$ neighbors of $v'$.
	Recall that $G$ is the subgraph of $G'$ induced by the set $\{v'\in V'_v \ |\  v \in V\}$. Then for each round $i\geq 0$ and for each   $v'\in \Active_{G'}[\bs,i]$,  we get that the set  $\Active_G[S,i]$ contains the corresponding vertex $v$.
Consequently $v$ will be activated in $G$. {
One can see that the same graph $G'$ can be used  to derive a similar reduction from TSS to WTSS.}
\qed

\begin{figure}	
	\includegraphics[height=4.3truecm]{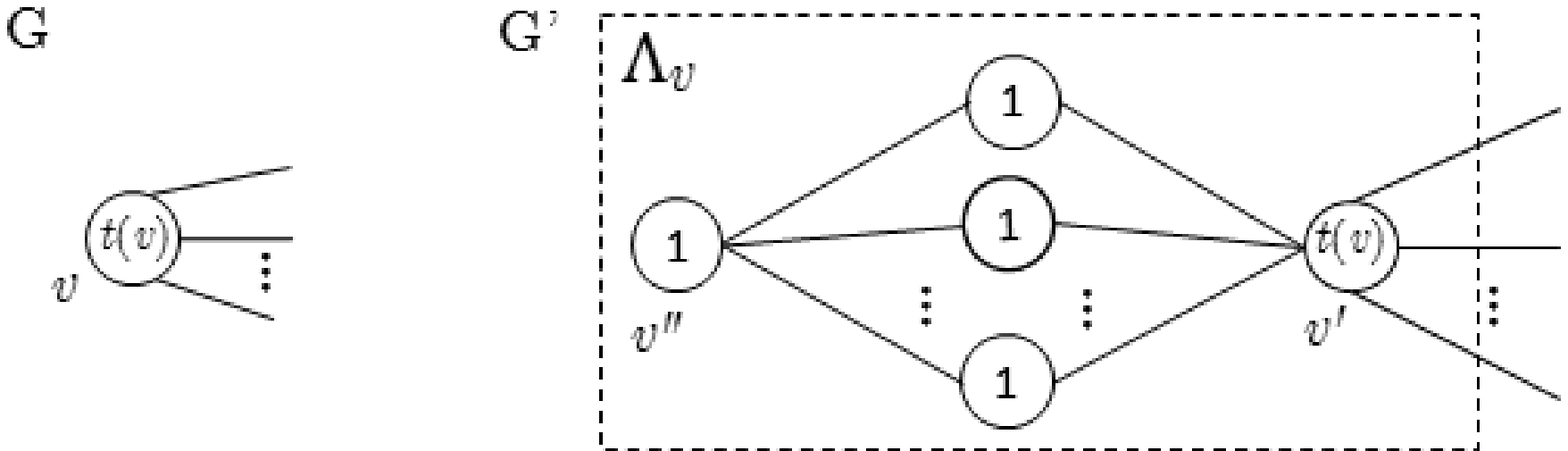}
		\caption{The gadget $\gad_v$: (left) a generic vertex $v \in V$ 
		having degree $d_G(v)$ and threshold $t(v)$; (right) the gadget $\gad_v$, having $d_G(v)+2$ vertices, associated to $v$.  \label{fig:gadget}}
\end{figure}

\section{The Algorithm for  Weighted Target Set Selection  }\label{sec:WTSS}  
Our  algorithm WTSS works by iteratively deleting vertices from the input graph $G$.
At each iteration, the vertex to be deleted is chosen as to maximize a certain
function (Case 3). During the deletion process, some vertex $v$ in the surviving graph
may remain with less neighbors than its threshold; in such a case (Case 2) $v$  is added
to the target set and deleted from the graph while its neighbors' thresholds are
decreased by $1$ (since they receive $v$'s influence). It can also happen that the surviving graph contains a vertex $v$ whose threshold has been decreased down to
$0$ (e.g., the deleted vertices are able to activate $v$); in such a case (Case 1) $v$
is deleted from the graph and its neighbors' thresholds are decreased by $1$ (since
as soon as vertex  $v$ activates, its neighbors will receive $v$'s influence).
\\
\begin{algorithm}
\SetCommentSty{footnotesize}
\SetKwInput{KwData}{Input}
\SetKwInput{KwResult}{Output}
\DontPrintSemicolon
\caption{ \ \   \textbf{Algorithm} WTSS($G$) \label{alg}}
\KwData { A graph $G=(V,E)$ with thresholds $t(v)$  and costs $c(v)$, for $v\in V$.\\ }
\KwResult{  A target set $S$ for $G$.}
\setcounter{AlgoLine}{0}
$S=\emptyset$;\\
$U=V$; \\  
\ForEach {$v\in V $}{
  $\d( v)=d_G(v)$;\\ 
  $k( v)=t( v)$; \\
  $N( v)=\Gamma_G( v)$; 
}
\While(\tcp*[f]{Select one vertex and eliminate it from the graph.}){ $U\neq \emptyset$ }{
    \eIf(\tcp*[f]{\underline{Case 1}: The  vertex  $v$ is activated by the influence of its neighbors in $V-U$ only; it can then influence its neighbors in $U$.}){there exists $v\in U$ s.t. $k(v)=0$}
    		{
    		\ForEach {$u\in N(v)$}{
    			$k(u)=\max(k(u)-1,0)$;
    		  }
    		}
				{
				\eIf(\tcp*[f]{\underline{Case 2}:  $v$ is added to $S$, since not enough  neighbors remain in $U$ to activate it;  $v$  can then influence its  neighbors in $U$.}){ there exists $v\in U$ s.t. $\d(v) <  k(v)$}
    				{$S=S\cup\{v\}$; \\
    				\ForEach {$u\in N(v)$}{
    						$k(u)=k(u)-1$;
    				}
						}
						{\tcp*[f]{\underline{Case 3:} The  vertex $v$ will  be  activated its  neighbors in $U$. }	\\
    				$v={\tt argmax}_{u\in U}\left\{\frac{c(u)\, k(u)}{\delta(u)(\delta(u)+1)}\right\}$;
    				}
						}
						\ForEach(\tcp*[f]{Remove the selected vertex $v$ from the graph.}) {$u\in N(v)$}{
    						$\d(u)=\d(u)-1$;\\
								$N(u)=N(u)-\{v\}$;
    				}
						$U=U-\{v\}$;				
}
\end{algorithm}	
\begin{example}\label{ex-2a}
Consider the tree $T$  in Figure \ref{figure2}.
The number inside each circle is the  vertex threshold and   $c(v)=t(v)$, for each $v$.
The algorithm  removes  vertices from $T$ 
 as in the table below  where,  
  for  each iteration of the while loop,   we give the  selected vertex and  which among Cases 1,  2  or 3 applies.
\begin{center}
\begin{tabular}{|l|l|l|l|l|l|l|l|l|l|l|}
\hline
Iteration  \quad & 1\quad & 2\quad & 3\quad & 4\quad & 5\quad & 6\quad & 7\quad & 8\quad & 9\quad & 10
\\
\hline
Vertex  \quad & $v_{5}$ \quad & $v_{10}$ \quad & $v_6$ \quad & $v_9$ \quad & $v_7$ \quad & $v_8$ \quad & $v_1$ \quad & $v_4$ \quad & $v_3$ \quad & $v_2$
\\
\hline
Case  \quad & 3\quad & 3\quad & 2\quad & 3\quad & 3\quad & 2\quad & 3\quad & 2\quad & 2\quad & 2\\
 \hline
\end{tabular}
\end{center}
The  set returned by the algorithm is 
$S=\{v_2, v_3, v_4, v_6,v_8\}$, a target set having  cost $\c(S)=5$.
\end{example}
\begin{figure}\label{figure2}
\begin{center}
\includegraphics[width=10truecm]{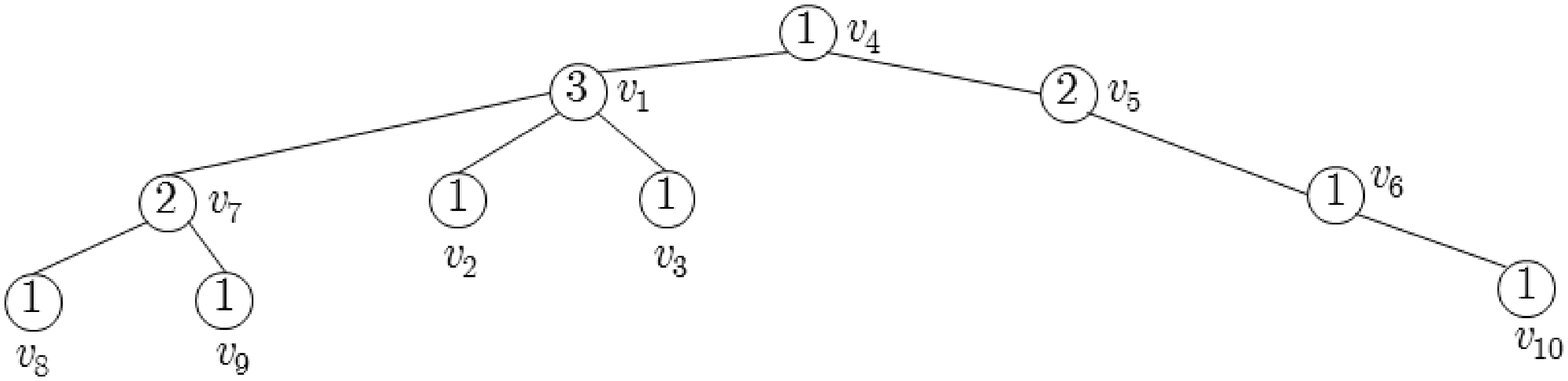}
\caption{The tree of Example \ref{ex-2a}. The number inside each circle rewpresents the vertex threshold.}
\end{center}
\end{figure}

\bigskip

The algorithm WTSS is a generalization to weighted graphs of the TSS algorithm presented in \cite{CGM+}.
The correctness of the algorithm  WTSS does not depend on the cost values, hence it immediately follows from the
correctness proof given in  \cite{CGM+}. Moreover a proof on the  bound on the target set size can be immediately obtained from the proof of the corresponding bound in \cite{CGM+}---by appropriately substituting the  threshold value $t(v)$ by the weighted  value $c(v)t(v)$ in the proof.
\begin{theorem}\label{teoCorr+teo2}
For any graph $G$ and threshold function $t$  the algorithm WTSS($G$) outputs a target set for $G$. The algorithm can be implemented so to run in time
$O(|E| \log |V |)$.
Moreover,  the algorithm WTSS($G$) returns a target set 
$S$ of cost
\begin{equation}\label{eq-soglia}
\c(S)\leq \sum_{v\in V} \frac{  c(v)t(v)}{d_G(v) +1}.
\end{equation}
\end{theorem}
\begin{theorem}\label{teoKint} 
The algorithm  
WTSS($G$) outputs an \emph{optimal target set}  if  $G$ is  a complete graph
  such that $c(v)\leq c(u)$  whenever $t(v)\leq t(u)$.
\end{theorem}
\proof
We denote by  $v_i$ the vertex selected during the $n-i+1$-th iteration of the while loop in the algorithm WTSS  and by $\G(i)$ the graph induced by the   vertices $v_{i},\ldots,v_1$, for $i=n,\ldots,1$.
	We show that for each $i=1,\ldots, n$ it holds that 
	 $S\cap \{v_i,\ldots,v_1\}$ is optimal for $\G(i)$.  
 Consider first $\G(1)$ consisting of the isolated vertex $v_1$ with threshold $k_1(v_1)$. 
It holds
$$\c(S \cap \{v_1\})=\begin{cases} {\c(\emptyset)=0}&{if \ k_1(v_1)=0}\\
                        {\c(\{v_1\})=c(v_1)}&{if \   k_1(v_1)>0} \end{cases}$$
												which is optimal.
Suppose now $\c(S\cap \{v_{i-1},\ldots,v_1\})$ is optimal for $\G(i-1)$ and consider $\G(i)$.
The selected vertex is $v_i$.
\\
If $k_i(v_i)=0$ then it is obvious that no optimal solution  for $\G(i)$ includes the ``already'' active vertex 
$v_i$. Hence, the inductive hypothesis on $\G(i-1)$ implies that     $\c(S\cap \{v_i,\ldots,v_1\})=\c(S\cap \{v_{i-1},\ldots,v_1\})$ is  optimal 
for $\G(i)$.
\\
If $k_i(v_i)>\d_i(v_i)=i-1$ then any optimal solution for $\G(i)$  includes vertex $v_i$ (which cannot be influenced otherwise) and the optimality follows by the optimality hypothesis on $\G(i-1)$.
\\
If none of the above holds, then  $0<k_i(v_j)\leq\d_i(v_j)$, for each $j\leq i$, and $c(v_i)k_i(v_i)\geq c(v_j)k_i(v_j)$, for each $j \leq i-1$.
We show now that there exists at least one optimal solution for $\G(i)$ which does not include $v_i$.
Consider an optimal solution $S^*_i$ for $\G(i)$ and assume $v_i\in S^*_i$. 
Let 
$$v= \argmax_{\myatop{1\leq j\leq i-1}{ v_j\notin S^*_i}} k_i(v_j). $$
By hypothesis the costs are ordered according  to the initial thresholds of the vertices.  
Since  at each step either all thresholds are decreased or they are all left equal,
we have that $c(v_j)k_i(v_j)\leq c(v_h)k_i(v_h)$ whenever $c(v_j)\leq c(v_h)$.
Hence, $\c(S^*_i-\{v_i\}\cup \{v\})\leq \c(S^*_i)$.
Moreover, recalling that $k_i(v_i)\leq\d_i(v_i)$ we know that $S^*_i-\{v_i\}\cup \{v\}$ is a solution for $\G(i)$.
\\
We have then found an optimal solution that does not contain $v_i$. This fact and the optimality
hypothesis on $\G(i-1)$ imply the optimality of $S\cap \{v_i,\ldots,v_1\}=S\cap \{v_{i-1},\ldots,v_1\}$.{\qed}

\bigskip

\section{Targeting with Partial Incentives} \label{sec:TPI}
In this section, we design an algorithm to efficiently allocate
incentives to the vertices of a network, in such a way that it triggers
an influence diffusion process that influences the whole network.
The algorithm is called TPI($G$). It   is close in spirit to the algorithm WTSS$(G)$, with some
crucial differences.
Again the algorithm proceeds by iteratively deleting vertices
from the graph and 
at each iteration 
the vertex to be deleted is chosen as to maximize a certain parameter (Case 2). 
If, during the deletion process,
a vertex $v$ in the surviving graph  remains with less neighbors 
than its remaining threshold   (Case 1), 
then $v$'s partial incentive is increased  
so that the $v$'s remaining threshold is at least as large as the number of $v$'s neighbors
in the surviving graph.
%

\newpage

\begin{algorithm}
\SetCommentSty{footnotesize}
\SetKwInput{KwData}{Input}
\SetKwInput{KwResult}{Output}
\DontPrintSemicolon
\caption{ \ \   \textbf{Algorithm} TPI($G$) \label{alg2}}
\KwData { A graph $G=(V,E)$ with thresholds $t(v)$, for each $v\in V$. }
\KwResult{ $\bs$ a target vector for $G$.}
\setcounter{AlgoLine}{0}
$U=V$; \\  
\ForEach {$v\in V $}{
	$\cs( v)=0;$  \tcp*[f]{Partial incentive initially assigned to $v$.} \\
  $\d( v)=d_G(v)$;\\ 
  $k( v)=t( v)$; \\
  $N( v)=\Gamma_G( v)$; 
}
\While(\tcp*[f]{Select one vertex and either update its incentive or remove  it from the graph.}){ $U\neq \emptyset$ }{
\eIf(\tcp*[f]{\underline{Case 1}:  Increase $\cs(v)$ and update $k(v)$}.){there exists $v\in U$ s.t. $k(v)>\d(v)$}
		{
		$\cs(v)=\cs(v)+ k(v)-\d(v)$;\\
		$k(v)=\d(v)$;\\
		\If(\tcp*[f]{here\  $\d(v)=0$.}){k(v)=0}{$U=U-\{v\};$}
		}
		(\tcp*[f]{\underline{Case 2}: Choose a vertex $v$ to eliminate from the graph. })	
				{
    		$v={\tt argmax}_{u\in U}\left\{\frac{k(u)(k(u)+1)}{\delta(u)(\delta(u)+1)}\right\}$;\\
				\ForEach{$u\in N(v)$}{
    						$\d(u)=\d(u)-1$;\\
								$N(u)=N(u)-\{v\};$
    				}
						$U=U-\{v\};$
			  }	
}
\end{algorithm}

\begin{example}\label{ex-4a}
Consider a complete graph on 7 vertices  
with thresholds $t(v_1)=\ldots= t(v_{5})=1$,   $t(v_{6})=t(v_{7})=6$ (cfr. Fig. \ref{fig:k}).
A possible execution of the algorithm is summarized below. 
At each iteration of the while loop, the algorithm   considers  the vertices  in the  order
shown in the table  below, where  we also 
 indicate for each vertex whether  Cases 1 or 2 applies and the updated value of the 
partial incentive for the selected vertex:

\begin{center}
\begin{tabular}{|l|l|l|l|l|l|l|l|l|}
\hline 
Iteration      & 1    & 2      & 3      & 4    & 5    & 6    & 7    & 8   \\
\hline 
vertex      & $v_{7}$       & $v_{6}$     & $v_{6}$    & $v_1$     & $v_2$        & $v_3$          & $v_4$     & $v_5$ \\
\hline 
Case      & 2    & 1        & 2      & 2         & 2      & 2       & 2    & 1\\
\hline 
 Incentive    & $0$     & $1$       & $1$   
                         & $0$      & $0$     & $0$       &  $0$     &  $1$\\
 \hline 
\end{tabular}
\end{center}
The algorithm  $TPI(G)$  outputs  the  vector of partial incentives having non zero elements
$s(v_5)=s(v_6)=1$, for which we have 
\begin{eqnarray*}
 \Active[\bs,0] &=& \{v_5\}  \qquad \mbox{\em (since $s(v_5)=1=t(v_5)$)} 
 \\
 \Active[\bs,1]&=&\Active[\bs,0]\cup\{v_1,v_2,v_3,v_4\}=\{v_1,v_2,v_3,v_4,v_5\}
 \\
 \Active[\bs,2]&=&\Active[\bs,1]\cup \{v_6   \} =\{v_1,v_2,v_3,v_4,v_5, v_6   \} \qquad \mbox{\em (since $s(v_6)=1$)}  
\\
 \Active[\bs,3]&=&\Active[\bs,2]\cup \{  v_7 \}   =V.
\end{eqnarray*}

\begin{figure}
\begin{center}
\includegraphics[width=6truecm]{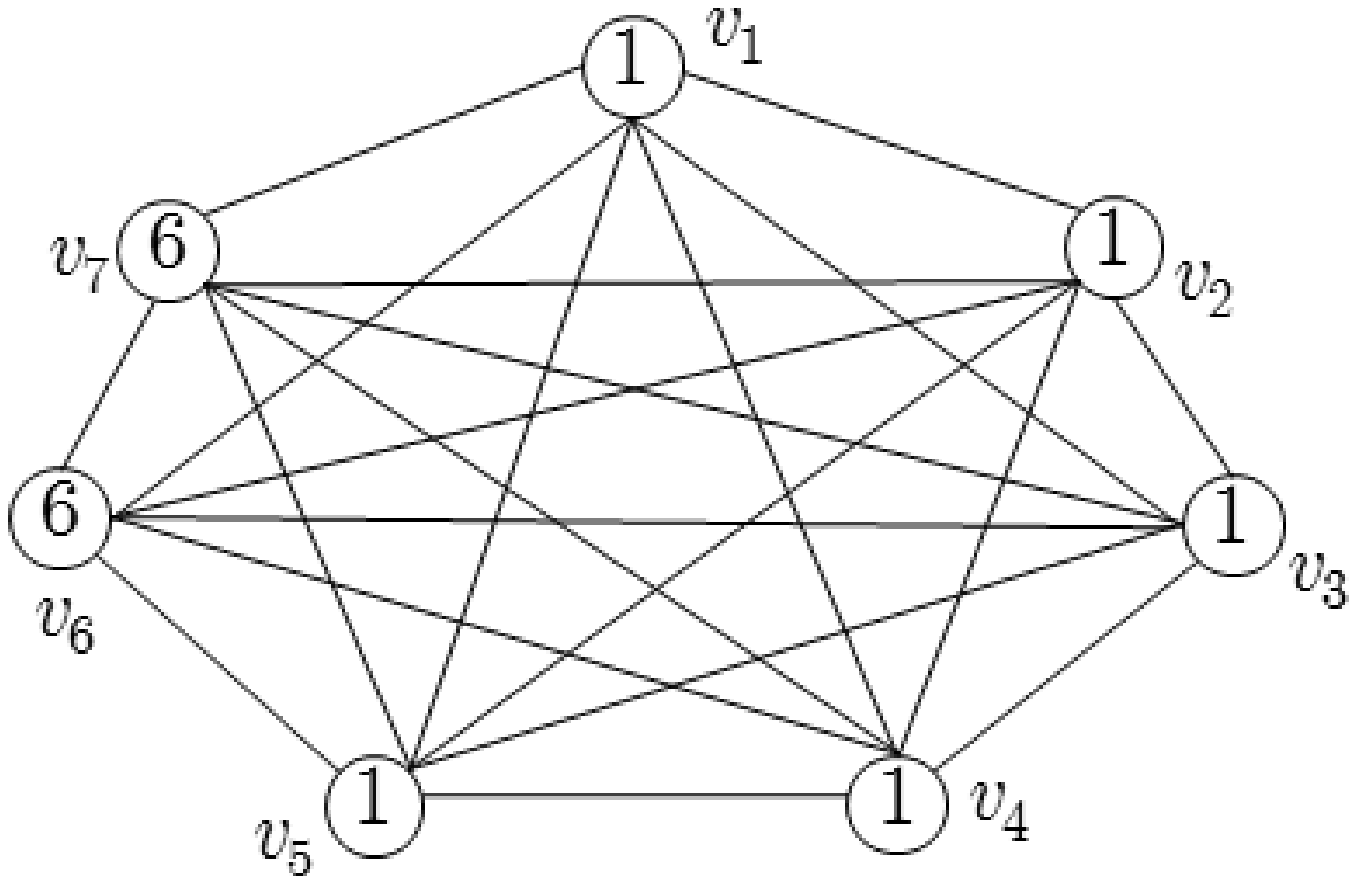}
\caption{A complete graph example. The number inside each circle is the  vertex threshold.\label{fig:k}}
\end{center}
\end{figure}

\end{example}

\noindent
We first prove the algorithm correctness, next we give a general  upper bound on the size 
$\sum_{v\in V}s(v)$ of
its output and prove its optimality for trees and cliques.

To this aim we   will use the following notation. 

Let $\ell$ be the number of iterations of the while loop in TPI($G$).
For each iteration $j$, with $1 \leq j \leq \ell$, of the while loop  we denote
\begin{itemize}
\item by $U_j$ the set $U$ 
at the beginning of the $j$-th iteration (cfr. line 7 of $TPI(G)$), in particular $U_1=V(G)$ and $U_{\ell+1}=\emptyset$;
\item by $\G(j)$  the subgraph of $G$ induced by the vertices in $U_j$,
\item by $\vj$ the vertex selected during the $j$-th iteration\footnote{A vertex can be  selected  several  times before being eliminated; indeed in Case 1  we can have $U_{j+1}=U_j$.}, 
\item by $\d_j(v)$  the degree of vertex $v$ in $\G(j)$, 
\item by  $k_j(v)$  the value of the remaining threshold of vertex $v$ in $\G(j)$, that is, as it is updated at 
the beginning of the $j$-th iteration, in particular $k_1(v)=t(v)$ for each $v\in V$,
\item by $\cs_j(v)$  the partial incentive collected by vertex $v$ in $\G(j)$ 
starting from  the  $j$-th iteration, in particular we set   $\cs_{\ell+1}(v)=0$ for each $v\in V$;    
\item by $\s_j$  the increment of  the partial incentives  during the $j$-th iteration, that  is,
$$\s_j=\cs_j(v_j)-\cs_{j+1}(v_j)=\begin{cases} {0}&{\mbox{if} \ k_j(\vj) \leq \d_j(\vj),}\\
                      {k_j(\vj) - \d_j(\vj)}&{\mbox{otherwise.}} 
				\end{cases}$$
\end{itemize}
According to  the above notation, we have that if vertex $v$ is  selected during the iterations
$j_1 < j_2 < \ldots < j_{a-1} < j_a$ of the while loop in TPI($G$),  where  the last value  $j_a$  is the iteration when  $v$ 
has been eliminated from the graph, then 
$$\cs_j(v)=\begin{cases}  {\s_{j_1}+\s_{j_2}+\ldots   + \s_{j_a}}&{\mbox{if} \  j \leq j_1,}\\
               {\s_{j_b} +  \s_{j_{b+1}}+\ldots  + \s_{j_a} } &{\mbox{if} \ j_{b-1} < j \leq j_b {\leq j_a},}\\
                       {0}&{\mbox{if} \ j > j_a.}
				\end{cases}$$
{In particular when $j=j_a$, it holds that $\cs_j(v)=\s_j$.}		

The following result is immediate.											
\begin{prop}\label{prop1}
Consider  the vertex $\vj$ that is  selected during   iteration $j$, for $1 \leq j \leq \ell$, of the while loop in the algorithm TPI($G$):
\begin{itemize}
\item[ {\bf 1.1)}] If Case 1 of the algorithm 
TPI($G$) holds and  $\d_{j}(\vj) =0$, then 
 $k_{j}(\vj)> \d_{j}(\vj) =0$ and  the isolated vertex $\vj$ is eliminated from $\G(j)$.
Moreover,\\  
$$U_{j+1}=U_{j} - \{\vj\},   \quad
\cs_{j+1}(\vj)= \cs_{j}(\vj)-\s_j,   \quad
 \s_{j}= k_{j}(\vj)-\d_{j}(\vj)=k_{j}(\vj)>0, $$ 
and, for each  $v {\in} U_{j+1}$
$$\cs_{j+1}(v)=\cs_{j}(v), \quad
 \d_{j+1}(v)= \d_{j}(v), \quad 
 k_{j+1}(v)= k_{j}(v).$$
\item[ {\bf  1.2)}] If Case 1 of  TPI($G$) holds with  $\d_{j}(\vj) >0$, then 
 $k_{j}(\vj)> \d_{j}(\vj) >0$ and  no vertex is deleted from $\G(j)$, that is, $U_{j+1}=U_{j}$.
 Moreover, 
$$\s_{j}= k_{j}(\vj)-\d_{j}(\vj)>0$$ 
and, for each $v \in U_{j+1}$
$$\d_{j+1}(v)= \d_{j}(v),\quad
\cs_{j+1}(v)=\begin{cases}  { \cs_{j}(\vj){-}\s_j }&{\mbox{if }  v=\vj}\\
													 {     \cs_{j}(v)        }&{\mbox{if } v \neq \vj }
													\end{cases}, \
\quad
 k_{j+1}(v)=\begin{cases}  {  \d_{j}(v)}&{\mbox{if} \ v=\vj}\\
													 {  k_{j}(v)   }&{\mbox{if} \  v \neq \vj}.
													\end{cases}												$$
\item[{\bf  2)}] If Case 2 of TPI($G$) holds then 
 $k_{j}(\vj) \leq \d_{j}(\vj)$ and 
$\vj$ is pruned from $\G(j)$. Hence, 
$$U_{j+1}=U_{j} - \{\vj\}, \qquad \s_{j}=0,$$
  and,  for each $v \in U_{j+1}$ it holds
$$
\cs_{j+1}(v)= \cs_{j}(v),  \quad  k_{j+1}(v)= k_{j}(v)		\quad 
\d_{j+1}(v)=\begin{cases}  {  \d_{j}(v) -1}&{\mbox{if} \ v \in \Gamma_{\G(j)}(\vj)}\\
													 { \d_{j}(v)   }&{\mbox{otherwise.} }
													\end{cases}
$$					
\end{itemize}
\end{prop}

\begin{lemma} \label{d=0}
For each iteration $j=1,2,\ldots,\ell$, of the while loop in the algorithm TPI($G$), 
\begin{itemize}
\item[1)] \ if  $\ k_j(\vj) > \d_j(\vj)$ then $\s_j= k_{j}(\vj)-\d_{j}(\vj)=1$;
\item[2)] \ if $\d_j(\vj)=0$ then $s_j(\vj)=k_j(\vj)$.
\end{itemize}
\end{lemma}
\noindent
{\proof
First, we prove 1).
At the beginning of the algorithm, $t(u)=k(u)\le d(u)=\d(u)$ holds for all  $u\in V$. 
Afterwords, the value of $\d(u)$ is decreased by at most one unit for each iteration 
(cfr. line 16 of \textit{TPI}($G$)). Moreover, 
the first time the condition of Case 1 holds for some vertex $u$, one has 
$\d_j(u)=k_j(u)-1$. Hence, if the selected vertex is $v_j=u$ then 1) holds; otherwise, some $v_j \neq u$, satisfying the condition of Case 1 is selected  and $\d_{j+1}(u)=\d_{j}(u)$ and $k_{j+1}(u)=k_{j}(u)$ hold.
Hence, when at some subsequent iteration $j'>j$ the algorithm selects $v_{j'} = u$, 
it holds  $\d_{j'}(u)=k_{j'}(u)-1$.
\\
To show  2), it is sufficient to notice that at the iteration $j$ when vertex $v_j$ 
is eliminated from the graph, it holds $\cs_j(v)=\s_j$.
}

\qed
\smallskip

Next theorem states the correctness of the algorithm TPI($G$) for any  graph $G$.
\begin{theorem}\label{teo-att-par}
For any  graph $G$ the algorithm TPI($G$) outputs a target vector for $G$.
\end{theorem}
\proof
We  show that for  each iteration $j$, with $1 \leq j \leq \ell$, the assignation of partial incentives
$s_j(v)$ for each $v \in U_j$ activates all the vertices of the graph $\G(j)$ when the distribution of  thresholds to its vertices is $k_j(\cdot)$.
The proof is by induction on $j$.
\\
If $j=\ell$ then  the unique vertex $v_\ell$ in $\G(\ell)$  has  degree $\d_\ell(v_\ell)=0$  and $s_\ell(v_\ell)=k_\ell(v_\ell)=1$ (see Lemma \ref{d=0}).
\\
Consider now $j< \ell$ and suppose the algorithm be correct on $\G(j+1)$ that is,  the assignation of partial incentives $s_{j+1}(v)$, for each $v \in U_{j+1}$, activates all the vertices of the graph $\G({j+1})$  when the distribution of thresholds to its vertices is $k_{j+1}(\cdot)$.  \\
Recall that $v_{j}$ denotes  the vertex the algorithm selects from  $U_j$ (thus  obtaining $U_{j+1}$,  the vertex set of $\G(j+1)$).
In order to prove the theorem we analyze  three cases according to the current degree and threshold of the selected vertex $v_{j}$.
\begin{itemize}
\item
Let $k_{j}(\vj)> \d_{j}(\vj) =0$.  By Lemma \ref{d=0},  we have $k_j(\vj)=s_j(\vj)$. Furthermore,  recalling that 1.1) of Proposition \ref{prop1} holds  and by using  the inductive hypothesis on $\G(j+1)$, we get the correctness on $\G(j)$ .
\item
Let $k_{j}(\vj)> \d_{j}(\vj) \geq 1$.  By recalling that  1.2) of Proposition \ref{prop1}  holds we get
 $k_j(v)- s_j(v)=k_{j+1}(v)-s_{j+1}(v)$, for each vertex $v \in U_j$. 
{Indeed, for each $v \neq v_j$ we have $k_{j+1}(v)=k_j(v)$ and $s_{j+1}(v)=s_{j}(v)$. 
Moreover, $$k_{j+1}(\vj)-s_{j+1}(\vj)= \d_j(\vj)-s_{j}(\vj)+\s_j=k_{j}(\vj)-s_{j}(\vj).$$}
   Hence the
vertices that can be activated in  $\G(j+1)$  can be activated in  $\G(j)$ with thresholds 
$k_{j}(\cdot)$ and partial incentives $s_{j}(\cdot)$. 
So, by using  the inductive hypothesis on $\G(j+1)$, we get the correctness on $\G(j)$.
\item
Let $k_{j}(\vj) \leq \d_{j}(\vj)$.
By recalling that  2) of Proposition \ref{prop1}  holds and by  the inductive hypothesis on  $\G(j+1)$ we have that all
the neighbors of $\vj$ in $\G(j)$ that are vertices in $U_{j+1}$ gets active; 
since $k_j(\vj) \leq \d_j(\vj)$ also $\vj$ activates in $\G(j)$.
\end{itemize}
\qed

\noindent
We now give an upper bound on the size of the solution returned by the algorithm TPI.
\begin{theorem}\label{teo3}
For any  graph  $G$ the algorithm TPI($G$) returns a target vector $\bs$ for $G$ such that
$$\Wbs=\sum_{v\in V}\cs(v)\leq \sum_{v\in V}  \frac{t(v)(t(v) +1)}{2(d_G(v)+1)}$$
\end{theorem}
\proof
Define  $\D(j)=\sum_{v \in U_j} \frac{k_j(v)(k_j(v)+1)}{2(\d_j(v)+1)}$,
for each $j=1,\ldots, \ell$. 
By definition of $\ell$, we have $\G(\ell+1)$ is the empty graph;  we then define $\D(\ell+1)= 0$.
We prove now by induction on $j$ that  
\begin{equation}\label{S}
\s_j \leq \D(j) - \D(j+1).
\end{equation}
By using (\ref{S}) we will have the bound on $\sum_{v \in V}\cs(v)$. Indeed, 
$$\sum_{v \in V}\cs(v) = \sum_{j=1}^{\ell}\s_j \leq \sum_{j=1}^{\ell} (\D(j) - \D(j+1)) 
=\D(1)-\D(\ell+1) = \D(1) =  \sum_{v\in V}  
\frac{t(v)(t(v) +1)}{2(d(v)+1)}.$$
\\
In order to prove (\ref{S}), we analyze  three cases depending on {the relation between $k_{j}(\vj)$ and $\d_{j}(\vj)$}.\\
$\bullet$ Assume first $k_{j}(\vj)> \d_{j}(\vj) =0$. We get
\begin{eqnarray*}
\D(j)- \D(j{+}1) &=& \sum_{v\in U_j} \frac{k_j(v)(k_j(v)+1)}{2(\delta_j(v)+1)}
     -\sum_{v\in U_{j+1}} \frac{k_{j+1}(v)(k_{j+1}(v)+1)}{2(\delta_{j+1}(v)+1)}
\nonumber\\
&=& \frac{k_j(\vj)(k_j(\vj)+1)}{2(\delta_j(v_j)+1)} + \sum_{v\in U_j -\{v_j\}} \frac{k_j(v)(k_j(v)+1)}{2(\delta_j(v)+1)}\\
     && \hphantom{aaaaaaaaaaaaaaa} - \sum_{v\in U_{j+1}} \frac{k_{j+1}(v)(k_{j+1}(v)+1)}{2(\delta_{j+1}(v)+1)}
\nonumber\\
&=& 
\frac{k_j(\vj)(k_j(\vj)+1)}{2(\delta_j(v_j)+1)}  \qquad \qquad\mbox{{(by 1.1 in Proposition \ref{prop1})}}\\ 
&=& 1 = \s_j. \qquad \qquad\qquad \qquad\qquad\mbox{{(by Lemma \ref{d=0})}}
\end{eqnarray*}

\noindent
$\bullet$ Let now $k_{j}(\vj)> \d_{j}(\vj) \geq 1$. We have
\begin{eqnarray*}
\D(j)- \D(j{+}1)&=& \sum_{v\in U_j} \frac{k_j(v)(k_j(v)+1)}{2(\delta_j(v)+1)}
     -\sum_{v\in U_{j+1}} \frac{k_{j+1}(v)(k_{j+1}(v)+1)}{2(\delta_{j+1}(v)+1)}
\nonumber\\
&=& \frac{k_j(\vj)(k_j(\vj){+}1)}{2(\delta_j(v_j)+1)} -\frac{k_{j+1}(\vj)(k_{j+1}(\vj){+}1)}{2(\delta_{j+1}(v_j)+1)}
\nonumber\\
&& +  \sum_{v\in U_j -\{v_j\}} \frac{k_j(v)(k_j(v)+1)}{2(\delta_j(v)+1)}
     -\sum_{v\in U_{j+1} -\{v_j\}} \frac{(k_{j+1}(v)(k_{j+1}(v)+1)}{2(\delta_{j+1}(v)+1)}
\nonumber\\
&=& 
\frac{(\delta_j(v_j)+1)(\delta_j(v_j)+2)}{2(\delta_j(v_j)+1)} - \frac{\delta_j(v_j)(\delta_j(v_j)+1)}{2(\delta_j(v_j)+1)}
\nonumber \ \  \mbox{{(by 1.2 in Proposition \ref{prop1})}}\\
&=& 
\frac{2(\delta_j(v_j)+1)}{2(\delta_j(v_j)+1)} = 1 = \s_j.  \qquad \qquad\qquad \qquad\qquad\mbox{{(by Lemma \ref{d=0})}}
\end{eqnarray*}

\noindent
$\bullet$ Finally, let $k_{j}(\vj)  \leq \d_{j}(\vj)$. 
In this case, by the algorithm  we know that   

\begin{equation} \label{argmaxEq}
\frac{k_j(v)(k_j(v)+1)}{\delta_j(v)(\delta_j(v)+1)} \leq \frac{k_j(\vj)(k_j(\vj)+1)}{\delta_j(\vj)(\delta_j(\vj)+1)},
\end{equation}
for each $v\in  U_j$.
Hence, we get
\begin{eqnarray*}
\D(j)-\D(j+1)  &=& \sum_{v\in U_j} \frac{k_j(v)(k_j(v)+1)}{2(\delta_j(v)+1)}
     -\sum_{v\in U_{j+1}} \frac{k_{j+1}(v)(k_{j+1}(v)+1)}{2(\delta_{j+1}(v)+1)} \\
&=& \frac{k_j(\vj)(k_j(\vj){+}1)}{2(\delta_j(v_j){+}1)} 
+  \sum_{v\in \Gamma_{\G(j)}(v_j)} \frac{k_j(v)(k_j(v){+}1)}{2(\delta_j(v){+}1)}\\
&& \hphantom{a}
     -\sum_{v\in \Gamma_{\G(j)}(v_j) } \frac{k_{j+1}(v)(k_{j+1}(v){+}1)}{2(\delta_{j+1}(v){+}1)} 
		               \qquad\qquad \mbox{(by 2 in Proposition \ref{prop1})}
\\
&=& 
 \frac{k_j(\vj)(k_j(\vj){+}1)}{2(\delta_j(v_j){+}1)} 
+  \sum_{v\in \Gamma_{\G(j)}(v_j)} \frac{ k_j(v)(k_j(v){+}1)}{2}\left( \frac{1}{(\delta_j(v)+1)}
     - \frac{1}{\delta_{j}(v)} \right)
\\
&=& 
 \frac{k_j(\vj)(k_j(\vj)+1)}{2(\delta_j(v_j)+1)} 
-  \sum_{v\in \Gamma_{\G(j)}(v_j)} \frac{k_j(v)(k_j(v)+1)}{2\delta_{j}(v)(\delta_j(v)+1)}
\\
&\geq & 
          \frac{k_j(\vj)(k_j(\vj)+1)}{2(\delta_j(v_j)+1)} 
-   \frac{k_j(v_j)(k_j(v_j)+1)\delta_{j}(v_j)}{2\delta_{j}(v_j)(\delta_j(v_j)+1)} \ \ \qquad	\mbox{(by (\ref{argmaxEq}))}\\ 
&=& 0 = \s_j
\qquad\qquad\qquad\qquad\qquad\qquad\qquad\qquad\qquad\qquad\qquad\qquad \Box
\end{eqnarray*}

\medskip
\noindent

\subsection{Complete graphs}
\begin{theorem}\label{teoKfrac} 
TPI($K$) returns an optimal  target vector  for any complete graph $K$.
\end{theorem}
\proof 
We will show that, for each  $j=\ell, \cdots,1$,  the  incentives $\cs_j(v)$ for $v\in U_j$ are optimal for  $\H(j)$ when the distribution of thresholds to its vertices is $k_{j}(\cdot)$.  In particular, we will prove that 
\begin{equation} \label{Sj}
\S_j = \sum_{h \geq j}\s_h =\sum_{v\in U_j}\cs_j(v) = \sum_{v\in U_j}\cs^*_j(v)
\end{equation}
for any optimal target vector $\bs^*_j$ for  $\H(j)$.

The theorem  follows by setting $j=1$ (recall that $\H(1)=K$ and  $\cs_1(v)=s(v)$, $k_1(v)=t(v)$  for each $v \in U_1 =  V(K)$).  
We proceed by induction on $j$.

For $j=\ell$, the graph  $\H(\ell)$ consists of the unique vertex $v_\ell$ and by Lemma \ref{d=0} and 1.1 of Proposition \ref{prop1}, it holds 
$\S_{\ell}=\s_{\ell}=\cs_{\ell}(v_\ell)=k_{\ell}(v_\ell)=1=\cs^*_{\ell}(v_\ell)$.

Consider now  some $j < \ell$ and suppose that  the partial incentives $\cs_{j+1}(v)$ for $v\in U_{j+1}$ are optimal for  $\H({j+1})$ when the distribution of thresholds  is $k_{j+1}(\cdot)$.
Consider the $j$-th iteration of  the while loop in TPI($K$).
First, notice that the complete graph $\H(j)$ cannot have isolated vertices; 
hence, only   1.2) and 2) in Proposition \ref{prop1} can hold for the selected vertex $\vj$.
We will prove that  (\ref{Sj}) holds. We distinguish two cases according to the value of the threshold $k_{j}(\vj)$.

\medskip
\noindent
Assume first that  $k_{j}(\vj)> \d_{j}(\vj)$. By 1.2) in Proposition \ref{prop1}  and the inductive hypothesis, we have 
$$\S_{j} = \sum_{h \geq j}\s_h 
          = \s_{j}+\sum_{h \geq j+1}\s_h
					= k_{j}(\vj)-\d_{j}(\vj) + \sum_{v\in U_{j+1}}\cs^*_{j+1}(v)  
					\leq \sum_{v\in U_{j}}\cs^*_{j}(v)$$
where the  inequality holds since any solution that optimally assigns  incentives $\bs^*_{j}$ to the vertices of $\H(j)$ increases by at least $k_{j}(\vj)-\d_{j}(\vj)$ the sum of the optimal partial incentives assigned to the vertices in $\H(j+1)$.

\medskip
\noindent
Suppose now that $k_{j}(\vj) \leq \d_{j}(\vj)$. By  2) in Proposition \ref{prop1}  and the inductive hypothesis we have
\begin{equation} \label{sum}
S_{j} = \sum_{h \geq j}\s_h 
          = \s_{j}+\sum_{h \geq j+1}\s_h
					=  0 + \sum_{v\in U_{j+1}}\cs^*_{j+1}(v).
\end{equation}
We will show that, given any optimal  incentive assignation $\cs^*_{j}(\cdot)$  to the vertices in $\H(j)$, it holds
\begin{equation} \label{ottimo}
\S_{j} \leq \sum_{v\in U_{j}}\cs^*_{j}(v).
\end{equation}
thus proving (\ref{Sj})  in this case.
Consider the activation process in $\H(j)$ that starts with the partial incentives $\cs^*_{j}(\cdot)$ and let $\tau$ be the round during which vertex $\vj$ is activated, that is
\begin{equation} \label{tau}
\left|\Active[\bs^*_{j},\tau-1] \cap \Gamma_{\H(j)}(\vj)\right| = k_{j}(\vj) - \cs^*_{j}(\vj).
\end{equation}
Equality  (\ref{tau}) implies that there exist 
$$\d_{j}(\vj)- (k_{j}(\vj) - \cs^*_{j}(\vj)) \geq \d_{j}(\vj)- \d_{j}(\vj)+\cs^*_{j}(\vj)= \cs^*_{j}(\vj)$$neighbors of $\vj$ in $\H(j)$ that will be activated in some round    larger or equal to $\tau$. 
Let $X$ be any subset of  $\cs^*_{j}(\vj)$ such neighbors (i.,e., $|X|=\cs^*_{j}(\vj)$) and define
\begin{equation} \label{z}
 z_{j}(v)=\begin{cases}  
{ \cs^*_{j}(v) +1}&{\mbox{if} \ v \in X,}\\
{ \cs^*_{j}(v)   }&{\mbox{if} \ v \in U_{j+1} -  \X,}\\
{ 0}&{\mbox{if} \ v=\vj.}
\end{cases}
\end{equation}
It is easy to see that the  incentives  $z_{j}(v)$ for $v \in U_{j}$ give a solution for 
$\H(j)$. 
Indeed, each vertex $v \in U_{j+1} - X$ activates at the same round as in the activation process 
starting with incentives  $\cs^*_{j}$; furthermore,
each vertex $v \in X$ can activate without the activation of $\vj$; finally, 
$\vj$ activates after  both vertices in $U_{j+1} -  X$ and vertices in $X$ are activated.\\
By the above and recalling 2) of Proposition \ref{prop1}, we  have that $z_{j}(v)$ for $v \in U_{j+1}$ is a solution for $\H(j+1)$. Hence, $\sum_{v\in U_{j+1}}z_{j}(v) \geq  \sum_{v\in U_{j+1}}\cs^*_{j+1}(v)$ and by (\ref{z}) and (\ref{sum})
we have
$$ \sum_{v\in U_{j}}\cs^*_{j}(v) = |X|+\sum_{v\in U_{j+1}}\cs^*_{j}(v)= \sum_{v\in U_{j+1}}z_{j}(v) \geq  \sum_{v\in U_{j+1}}\cs^*_{j+1}(v) = S_{j}$$
thus proving (\ref{ottimo}).
\qed

\bigskip
\subsection{Trees}
\noindent
In this section
 we prove  the optimality of the algorithm TPI when the input graph is a tree.
\begin{theorem}\label{teoTfrac} 
 TPI($T$) outputs an optimal target vector  for any tree $T$.
\end{theorem}
\proof 
 We will show, for each  $j=\ell, \cdots,1$, that the  incentives $\cs_j(v)$ for $v\in U_j$ are optimal for the forest  $\T(j)$ with   thresholds  $k_{j}(\cdot)$.  In particular, we will prove that 
\begin{equation} \label{Sjtree}
\S_j = \sum_{h \geq j}\s_h =\sum_{v\in U_j}\cs_j(v) = \sum_{v\in U_j}\cs^*_j(v)
\end{equation}
for any optimal target vector $\bs^*_j$ for  the vertices in $U_j =V(\T(j))$.

The theorem will follow for $j=1$ (recall that $\T(1)=T$ and  $\cs_1(v)=s(v)$, $k_1(v)=t(v)$  for each $v \in U_1 =  V(T)$).  
We proceed by induction on $j$.

For $j=\ell$, the graph  $\T(\ell)$ consists of the unique vertex $v_\ell$ and by Lemma \ref{d=0} and  1.1) in Proposition \ref{prop1}, it holds $\S_{\ell}=\s_{\ell}=\cs_{\ell}(v_\ell)=k_{\ell}(v_\ell)=1=\cs^*_{\ell}(v_\ell)$.

Suppose now the partial incentives $\cs_{j+1}(v)$ for $v\in U_{j+1}$ are optimal for the forest $\T({j+1})$ when the  thresholds are $k_{j+1}(\cdot)$,  for some $j<\ell$.

Consider the $j$-th iteration of  the while loop in TPI($T$).
We will prove that  (\ref{Sjtree}) holds.
 We distinguish three cases according to the value of the  $k_{j}(\vj)$ and $\d_{j}(\vj)$.

\medskip
\noindent
Let $k_{j}(\vj)> \d_{j}(\vj)=0$. In such a case $v_j$ is an isolated vertex. By Lemma \ref{d=0},   1.1) of Proposition \ref{prop1},  and the inductive hypothesis we have
 $$\S_{j}=\sum_{h \geq j}\s_h 
          = \s_{j}+\sum_{h \geq j+1}\s_h
					=  k_{j}(\vj) + \sum_{v\in U_{j+1}}\cs^*_{j+1}(v)
					=  1 + \sum_{v\in U_{j+1}}\cs^*_{j+1}(v)
					\leq \sum_{v\in U_{j}}\cs^*_{j}(v).$$

\medskip
\noindent
Let $k_{j}(\vj)> \d_{j}(\vj)>0$.  By  1.2) in Proposition \ref{prop1} and the inductive hypothesis we have 
$$\S_{j} = \sum_{h \geq j}\s_h 
          = \s_{j}+\sum_{h \geq j+1}\s_h
					=  k_{j}(\vj)-\d_{j}(\vj) + \sum_{v\in U_{j+1}}\cs^*_{j+1}(v)
					\leq \sum_{v\in U_{j}}\cs^*_{j}(v),$$
where the  inequality follows since any solution that optimally assigns partial incentives $\bs^*_{j}$ to the vertices in $\T(j)$ increases of at least $k_{j}(\vj)-\d_{j}(\vj)$ the sum of the optimal  incentives assigned to the vertices in $\T(j+1)$.

\medskip
\noindent
Let  $k_{j}(\vj) \leq \d_{j}(\vj)$. By 2) in Proposition \ref{prop1}  and the inductive hypothesis we have
$$
S_{j} = \sum_{h \geq j}\s_h 
          = \s_{j}+\sum_{h \geq j+1}\s_h
					=  0 + \sum_{v\in U_{j+1}}\cs^*_{j+1}(v)
$$
In order to complete the proof in this case we will show that, given any optimal partial incentive assignment $\cs^*_{j}(\cdot)$  to the vertices in $\T(j)$, there is a cost equivalent optimal partial incentive assignment $z_{j}(\cdot)$ where $z_{j}(v_j)=0$. Moreover, this solution activates  also all the vertices in $\T(j+1)$. Hence
\begin{equation} \label{ottimoT}
\S_{j} = \sum_{v\in U_{j}}\cs^*_{j}(v).
\end{equation}
thus proving (\ref{Sjtree})  in this case.
\\
First of all we show that $k_{j}(\vj) = \d_{j}(\vj)$.
Indeed, for each leaf $u \in U_j$ we have $k_{j}(u) = \d_{j}(u)=1$, which maximizes the value 
$\frac{k_{j}(u)(k_{j}(u)+1)}{\d_{j}(u)(\d_{j}(u)+1)}$ since for any other vertex 
$v \in U_j$, $k_{j}(v) \leq \d_{j}(v)$. 
Hence,  $v_j$ is either a leaf vertex or an internal vertex with $k_{j}(\vj) = \d_{j}(\vj)$.
\\
Let $\Gamma_j(v_j)=\{u_1,u_2,\ldots,u_{\d_{j}(\vj)}\}$   be the set of $v_j$'s neighbors.
We have two cases to consider according to the value of  $\cs^*_{j}(v_j)$ 
\begin{itemize}
\item if $\cs^*_{j}(v_j)=0$, then we have $\cs^*_{j}(\cdot)=z_{j}(\cdot)$. Since  $k_{j}(\vj) = \d_{j}(\vj)$ and $\cs^*_{j}(v_j)=0$, each vertex in $\Gamma_j(v_j)$ is activated without the influence of $v_j$. 
Therefore,  $\cs^*_{j}(\cdot)$ is also a solution for $\T(j+1)$.
\item if $\cs^*_{j}(v_j)>0$, then we can partition  $\Gamma_j(v_j)$ into two sets: $\Gamma'_j(v_j)$ and 
$\Gamma''_j(v_j) $:
\begin{itemize}
 \item  $\Gamma'_j(v_j)$  includes $\d_{j}(\vj)-\cs^*_{j}(v_j)$ vertices that are activated before $v_j$ (this set must exist otherwise $v_j$  will never activate); 
\item $\Gamma''_j(v_j)\subseteq \Gamma_j(v_j)$ which consists of  the remaining $\cs^*_{j}(v_j)$ vertices. 
\end{itemize}
We define $z_{j}(\cdot)$ as follows:
$$ 
z_{j}(v)= \begin{cases}  0 & \mbox { if } v=v_j;	\\
																	\cs^*_{j}(v)+1 & \mbox { if } v \in \Gamma''_j(v_j); \\
																	\cs^*_{j}(v)  & \mbox { otherwise.} 
\end{cases}
$$
By construction we have that $\sum_{v \in U_j} z_{j}(v)=\sum_{v \in U_j} \cs^*_{j}(v)$.
 Moreover $z_{j}(v)$ activates all the vertices in $\T(j)$.
In particular, the vertices in $\Gamma'_j(v_j)$ activate  before $v_j$, while the vertices in  $\Gamma''_j(v_j)$ activate independently of $v_j$ thanks to the  increased incentive. 
Therefore $v_j$ activates thanks to the vertices in $\Gamma_j(v_j)$. A similar reasoning shows that $z_{j}(v)$ activates all the vertices in $\T(j+1)$.
\end{itemize}
\qed

We can also explicitly evaluate  the cost of an optimal solution for any tree.

\begin{theorem} \label{th:treeCost}

Any optimal target vector $\bs^*$ on a tree $T$  with thresholds $t: V \longrightarrow \N$ has cost 
\begin{equation}\label{eqT}
\c(\bs^*) = \sum_{v\in V} s^*(v) = |V|-1-\sum_{v \in V} d(v)-t(v).
\end{equation}
\end{theorem}
\proof
We proceed by structural induction on  $T$. If  $T$ consists of a single vertex $r$, then the optimal solution clearly has  $s^*(r)=t(r)$. Hence,  $\c(\bs^*)=s^*(r)=t(r)$ and \ref{eqT}) holds.
\\
Let now $T$ be a tree, with at least two vertices, rooted in   $r$.  
Let $\bs^*$ be an optimal target vector for $T$. 
The optimality of  $\bs^*$  clearly implies that $s^*(r)\leq t(r)$. 
Therefore, the root  $r$ needs to be influenced by  $t(r)-s^*(r)\geq 0$ of its children. Once $r$ is activated, it can influence  the  remaining children.
%
Summarizing, we have that there exists  
an ordering  $v_1,v_2, \ldots, v_{d(r)}$ of $r$'s children such that,
\begin{equation}\label{eqT2}
\c(\bs^*) = s^*(r)+\sum_{i=1}^{d(r)} \c(\bs^*_i),
\end{equation}
where $ \bs^*_i$ is an optimal target vector for the subtree $T(v_i)$ rooted at $v_i,$  assuming that  each vertex $v$ in $T(v_i)$ has  threshold  $t_i(v)$ given by
$$
t_i(v)=\begin{cases} 
t(v) & \mbox{if ($v\neq v_i$ for $i=1,\ldots, d(r)$) or}\mbox{ ($v=v_i$ for some  $1\leq i  \leq  t(r)-s^*(r)$)} \\
t(v_i)-1 & \mbox{if $v=v_i$ for some   $ t(r)-s^*(r)+1\leq i \leq d(r)$}.
\end{cases}
$$
 Let  $V_i$ denote  the  vertex set of    $T(v_i)$ and  $d_i(v)$ denote the  degree 
of   $v$ in $T(v_i)$---trivially, 
$d_i(v_i)=d(v_i)-1$ and $d_i(v)=d(v)$ for each $v\neq v_i$.

Assuming   by induction  that (\ref{eqT}) holds for    $T(v_i)$, for $i=1,\ldots,d(r)$,
by (\ref{eqT2}) we have
\begin{eqnarray*}
\c(\bs^*) 
& = & 
\bs^*(r)+\sum_{i=1}^{d(r)}\left(|V_i|-1-\sum_{v\in V_i}d_i(v){-}t_i(v)\right) \\
\\
& = & s^*(r) +\sum_{i=1}^{t(r)-s^*(r)}
       \left( |V_i|-1
			-\sum_{v\in V_i \atop v\neq v_i} (d(v){-}t(v)) \  
			 -\ (d(v_i)-1)  + t(v_i)  \right)  \\
& & 
 \qquad\  +\sum_{i=t(r)-s^*(r)+1}^{d(r)}
\left( |V_i|-1
-\sum_{v\in V_i\atop v\neq v_i}(d(v){-}t(v))\ 
                                        -(d(v_i)-1)  + (t(v_i)-1)  \right)   \\ 
																				\\
& = & s^*(r) +\sum_{i=1}^{t(r)-s^*(r)}
       \left( |V_i| -\sum_{v\in V_i \atop } (d(v){-}t(v))  \right)    
			          +\sum_{i=t(r)-s^*(r)+1}^{d(r)} \left( |V_i| -\sum_{v\in V_i}(d(v){-}t(v)) -1 \right)   \\ 
																				\\
& = & s^*(r) + \sum_{i=1}^{d(r)}|V_i|    -\sum_{i=1}^{d(r)} \sum_{v \in V_i} (d(v){-}t(v) )  
                                                 - (d(r) - t(r) +s^*(r)) \\
\\
&= & \left(|V|-1\right) -\sum_{v \in V } d(v){-}t(v) \qquad\qquad\qquad\qquad\qquad\qquad\qquad\qquad\qquad\qquad\qquad\quad\Box
\end{eqnarray*}
%

\bigskip

\section{Experiments}\label{sec:experiments} 
We have experimentally evaluated both our algorithms WTSS($G$) and TPI($G$) on real-world data sets and found that they perform quite  satisfactorily.
We conducted experiments on several real networks of various sizes from the Stanford
Large Network Data set Collection (SNAP) \cite{snap}, the Social Computing Data
Repository at Arizona State University \cite{ZL09} and  Newman's Network data  \cite{N15}. The data sets we considered include both
networks for which ``low cost'' target sets exist and networks needing an expensive target sets (due
to a community structure that appears to block the diffusion  process). 

\medskip

\subsection{Test settings}
\noindent{\em The competing algorithms.} We compare the
performance of our algorithms toward that of the best, to our knowledge, computationally
feasible algorithms in the literature \cite{Dem14}.
It is worth  mentioning that the following competing algorithms were initially designed  for the Maximally Influencing Set problem, where the goal is to identify a set $S \subseteq V$ such that its cost is bounded by a certain budget $\beta$ and the
activation process activates as much vertices as possible. In order to compare such algorithms toward our strategies, for each algorithm we performed a binary search in order to find the smallest value of $\beta$ which allow to activate all the vertices of the considered graph.
 We compare the WTSS algorithm toward the following two algorithms:
\begin{itemize}
\item \textit{DegreeInt,} a simple greedy algorithm, which selects vertices in descending order of degree \cite{KKT-03,CWY09};
\item \textit{DiscountInt,} a variant of DegreeInt, which selects a vertex $v$ with the highest degree at each step. Then the degree of vertices in $\Gamma(v)$ is decreased by $1$ \cite{CWY09}.
\end{itemize}
\noindent
Moreover, we compare the TPI algorithm toward the following two algorithms:
\begin{itemize}
	\item \textit{DegreeFrac}, which selects each vertex fractionally proportional to its degree. Specifically, given a graph $G=(V,E)$ and budget $\beta$ this algorithm spend on each vertex $v\in V,$ $s(v)=\left\lfloor\frac{d(v) \times  \beta}{2|E|}\right\rfloor$\cite{Dem14}. Remaining budget, if any, is assigned increasing by $1$ the budget assigned to some vertices (in descending order of degree). 
	\item \textit{DiscountFrac,} which at each step, selects the vertex $v$ having the highest degree and assigns to it a budged $s(v)=max(0,t(v)-|\Gamma(v)\cap S|))$, which represent the minimum amount that allows to activate $v$ 
	($S$ denotes the set of already selected vertices). As for the DiscountInt algorithm, after selecting a vertex $v$, the degree of vertices in $\Gamma(v)$ is decreased by $1$ \cite{Dem14}.
\end{itemize}

\bigskip

\noindent{\em Test Networks.} The main characteristics of the studied networks are shown in Table \ref{net}. In particular, for each network we report the number of vertices, the number of edges, the maximum degree, the diameter, the size of the largest connected component, the number of triangles, the clustering coefficient and the network modularity.
\begin{table}[ht]

\begin{center}
\resizebox{\linewidth}{!} {
\begin{tabular}{|l|r|r|r|r|r|r|r|r|}
\hline
Name 												& \# of vert. & \# of edges & Max deg & Diam.  & Size of the LCC & Triangles & Clust Coeff & Modul. \\ \hline
Amazon0302 \cite{snap}    	& 262111   	&  1234877  	  &  420      &     32    &     262111      &    717719 &  0.4198          &      0.6697      \\ \hline
BlogCatalog \cite{ZL09}     &   88784   &  4186390      &  9444     &     --     &     88784       &  51193389 &  0.4578          &      0.3182	\\ \hline
BlogCatalog2 \cite{ZL09}    &   97884   & 2043701 		  &    27849  &   5       &      97884      &   40662527&  0.6857          &      0.3282       \\ \hline
BlogCatalog3 \cite{ZL09}    &   10312   & 333983				&    3992   &   5       &      10312      &   5608664 &  0.4756          &      0.2374      \\ \hline
BuzzNet \cite{ZL09} 				&   101168  & 4284534 			&    64289  &   --       &      101163     &   30919848&  0.2508          &      0.3161     \\ \hline
ca-AstroPh \cite{snap} 			&   18772   & 198110 				&    504   	&   14      &      17903      &   1351441 &  0.6768          &      0.3072      \\ \hline
ca-CondMath \cite{snap} 		&   23133   & 93497 				&    279   	&   14      &      21363      &   173361  &  0.7058          &      0.5809      \\ \hline
ca-GrQc \cite{snap} 				&   5242    & 14496 				&    81    	&   17      &      4158       &   48260   &  0.6865          &      0.7433     \\ \hline
ca-HepPh \cite{snap} 				&   10008   & 118521 				&    491    &   13      &      11204      &   3358499 &  0.6115          &      0.5085     \\ \hline
ca-HepTh \cite{snap}				&   9877    & 25998 				&    65    	&   17      &      8638       &   28399   &  0.5994          &      0.6128     \\ \hline
Douban \cite{ZL09}					&   154907  & 327162 				&    287    &   9       &     154908      &   40612   &  0.048           &      0.5773     \\ \hline
Facebook \cite{snap} 				&  4039     & 88234 				&    1045   &   8       &      4039       &   1612010 &  0.6055          &      0.8093     \\ \hline
Flikr \cite{ZL09}		 				& 80513    & 5899822   	  	&    5706   &   --      &      80513      & 271601126 &  0.1652          &      --          \\ \hline
Hep \cite{snap} 						&  27770    & 352807 				&    64   	&   13      &     24700       &   1478735 &  0.3120          &      0.7203      \\ \hline
Last.fm \cite{ZL09}	 				& 1191812   & 5115300   		&    5140		&   --       &      1191805    &   3946212 &    0.1378        &       --    \\ \hline
Livemocha \cite{ZL09}				& 104438    & 2196188   		&    2980		&   6       &      104103     &   336651  &  0.0582          &      0.36     \\ \hline
Power grid     \cite{N15}   & 4941		 	&6594			   		&    19  		&   46      &      4941       &   651     &  0.1065          &      0.9306     \\ \hline
Youtube2 \cite{ZL09}				& 1138499 	&2990443	   		&    28754  &   --       &      1134890    &   3056537 &  0.1723          &       0.6506     \\ \hline
\end{tabular}
}
\end{center}
\caption{Networks parameters.  	\label{net}} 
\end{table}

\medskip

\bigskip

\noindent{\em Thresholds values.} We tested with three categories of  threshold function:
\begin{itemize}
\item \textit{Random thresholds} where $t(v)$ is chosen uniformly at random in the interval $[1,d(v)]$;
\item \textit{Constant thresholds} where the thresholds are constant among all vertices (precisely the constant value is an integer in the interval $[2,10]$ and for each vertex $v$ the threshold $t(v)$ is set as $\min(t,d(v))$ where $t=2,3, \ldots, 10$ (nine tests overall); 
\item \textit{Proportional thresholds} where for each  $v$ the threshold $t(v)$ is set as $\alpha \times d(v)$ with $\alpha=0.1,0.2, \ldots, 0.9$ (nine tests overall). Notice that for $\alpha=0.5$ we are considering a particular version of the activation process named ``majority'' \cite{FKRRS-2003}.
\end{itemize}

\bigskip

\noindent{\bf Costs.} We report experiments results for the WTSS problem in case the costs are equal to the thresholds, that is $c(v)=t(v)$ for each vertex $v\in V$. Similar results hold for different cost choices.

\begin{table}[ht]
\begin{center}
\resizebox{\linewidth}{!} {
\begin{tabular}{|l|r|r|r|r|r|r|r|}
\hline
& \multicolumn{3}{c|}{Targeting with Partial Incentives} &  &\multicolumn{3}{c|}{ Weighted Target Set Selection with $c(\cdot)=t(\cdot)$}\\ \hline
Name 		& PTI &	DiscountFrac	& DegreeFrac & &		WTSS &	DiscountInt	&DegreeInt  \\ \hline
Amazon0302	& 52703  	&328519 (623\%) & 	879624 (1669\%)	& & 	85410   &	596299 (698\%)	& 890347 (1042\%)\\ \hline
BlogCatalog	& 21761  &	824063 (3787\%) &	980670 (4507\%)	 & &	82502  	&1799719 (2181\%)	&2066014 (2504\%)\\ \hline
BlogCatalog2 &	16979   &	703383 (4143\%)	& 178447 (1051\%)	& &	67066  	&1095580 (1634\%)	&1214818 (1811\%)\\ \hline
BlogCatalog3 &	161  	&3890 (2416\%)&	3113 (1934\%)	& &	3925  &	3890 (99\%)&	3890 (99\%)\\ \hline
BuzzNet&	50913  	&1154952 (2268\%)&	371355 (729\%)	& &	166085  &	1838430 (1107\%)&	2580176 (1554\%)\\ \hline
ca-AstroPh &	4520  &	67189 (1486\%)&	198195 (4385\%)	& &	13242  &	183121 (1383\%)&	198195 (1497\%)\\ \hline
ca-CondMath &	5694  	& 31968 (561\%)	& 94288 (1656\%)	& &	10596  	&76501 (722\%)&	94126 (888\%)\\ \hline
ca-GrQc &	1422  	& 5076 (357\%)	& 15019 (1056\%)	& &	2141  	&12538 (586\%)&	15019 (701\%)\\ \hline
ca-HepPh &	4166  &	42029 (1009\%)&	120324 (2888\%)	& &	11338  &	118767 (1048\%)&	120324 (1061\%)\\ \hline
ca-HepTh &	2156  &	9214 (427\%)	&26781 (1242\%)	& &	3473  &	25417 (732\%)	&26781 (771\%)\\ \hline
Douban &	51167  &	140676 (275\%) &	345036 (674\%)	& &	91342  &	194186 (213\%)	&252739 (277\%)\\ \hline
Facebook &	1658  &	29605 (1786\%)&	54508 (3288\%)	& &	5531  &	77312 (1398\%)&	86925 (1572\%)\\ \hline
Flikr	& 31392  &	2057877 (6555\%)	& 134017 (427\%) & &	110227  &	5359377 (4862\%)	&5879532 (5334\%)\\ \hline
Hep	 & 4122  &	11770 (286\%) &	33373 (810\%)	& &	5526  	&33211 (601\%)&	33373 (604\%)\\ \hline
LastFM	& 296083  &	1965839 (664\%) &	4267035 (1441\%)	& &	631681  &	2681610 (425\%)&	4050280 (641\%)\\ \hline
Livemocha	& 26610  &	861053 (3236\%) &	459777 (1728\%)	& &	57293  &	1799468 (3141\%)	&2189760 (3822\%)\\ \hline
Power grid	& 767  &	2591 (338\%)&	4969 (648\%)	& &	974  &	3433 (352\%)&	4350 (447\%)\\ \hline
Youtube2	& 313786  &	1210830 (386\%)&	3298376 (1051\%)	& &	576482  &	2159948 (375\%)	& 3285525 (570\%)\\ \hline
\end{tabular} 
}
\end{center}
\caption{Random Threshold Results. 	\label{randomtest} }
\end{table}

\subsection{Results}
We compare the cost of the target set (or target vector) generated by six  algorithms (PTI, DiscountFrac, DegreeFrac, WTSS, DiscountInt, DegreeInt) on $18$ networks, 
fixing the thresholds in $19$ different ways (Random, Constant with $t=2,3, \ldots, 10$ and Proportional with $\alpha=0.1,0.2, \ldots, 0.9$). Overall we performed $6 \times 18 \times 19=2052$ tests.

\medskip 

\medskip
\noindent{\em Random Thresholds.}

Table \ref{randomtest} gives the results of the Random threshold test setting. 
Each number represents the cost of the target vector (left side of the table) or the target set (right side of the table) generated by each algorithm on each network using random thresholds (the same thresholds values have been used for all the algorithms). The value in bracket represents the overhead percentage compared to our algorithms (TPI for DiscountFrac	and  DegreeFrac  and WTSS for DiscountInt	and DegreeInt).
 Analyzing the results Table \ref{randomtest}, we notice that
in all the considered cases, with the exception of the network BlogCatolog3, our algorithms always outperform
their competitors. In the network BlogCatalog3, the WTSS algorithm is slightly worse than its competitors but PTI performs  much  better than the other algorithms.

\medskip

\medskip
\noindent{\em Constant and Proportional thresholds.}
The following figures depict the results of Constant and Proportional thresholds settings. For each network the results are reported in two separated figures:  Proportional thresholds (left-side), the
value of the $\alpha$ parameter appears along the $X$-axis, while the cost of the solution appears along the $Y$-axis;  Constant thresholds (right-side), in this case the $X$-axis indicates the value of the thresholds. 
{ We present the results only for {eight} networks;  the   experiments performed}
 on  the other networks  exhibit similar behaviors.
Analyzing the results from Figures  \ref{Amazon}-\ref{Amazon2}, we can make the following observations: In all the considered case our algorithms always outperform their competitors; the only algorithm that provides performance close to our algorithms is the DiscountFrac algorithm. However,  for intermediate values of the $\alpha$ parameter, the gap 
to our advantage is  quite significant. In general, in case of  partial incentives we have even  better results, the gap to our advantage increases with the increase of the parameter $\alpha$.

\begin{figure}%
\begin{center}
\includegraphics[width=0.50\linewidth, height =6.3truecm]{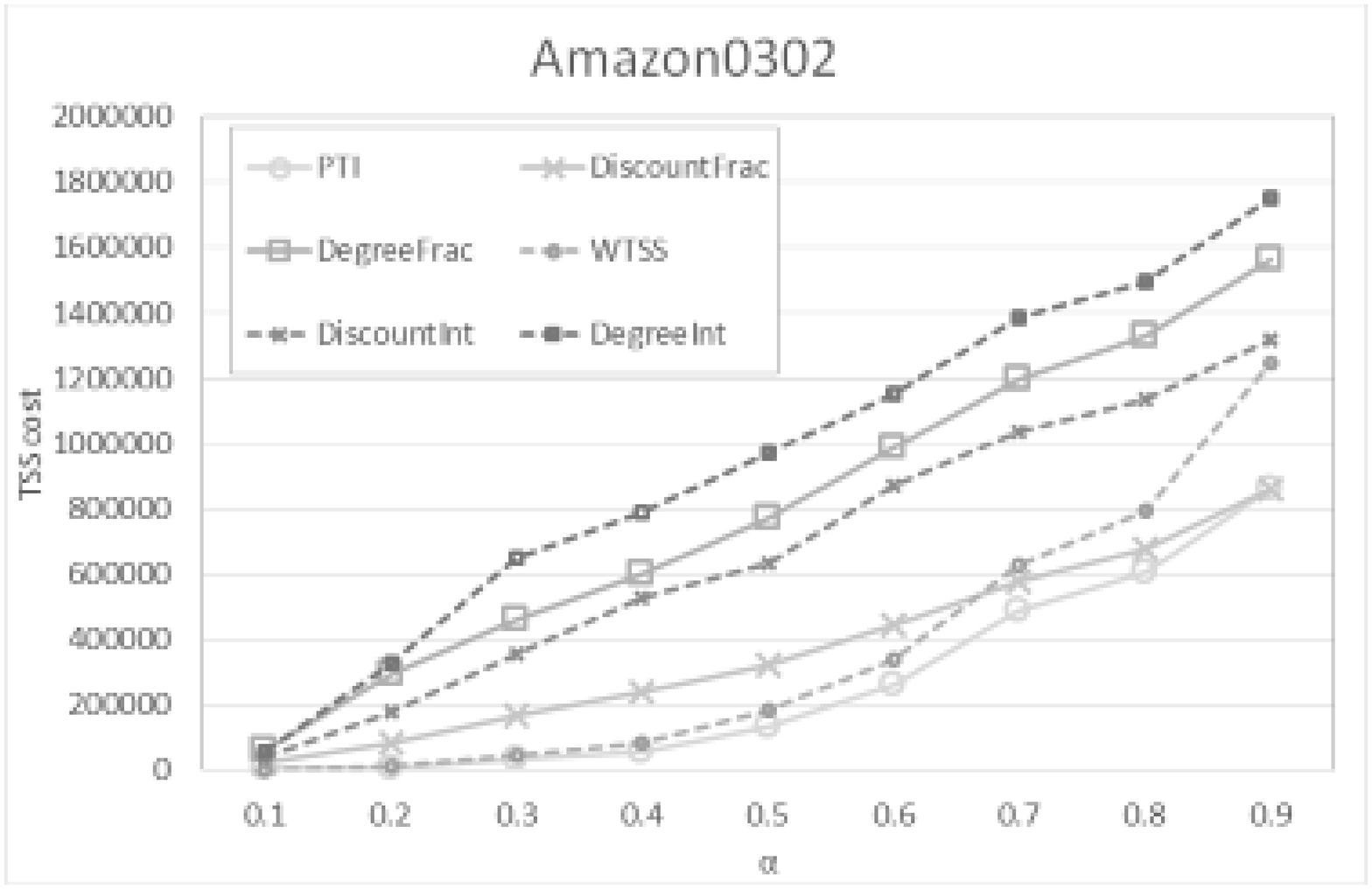}%
\includegraphics[width=0.50\linewidth, height =6.3truecm]{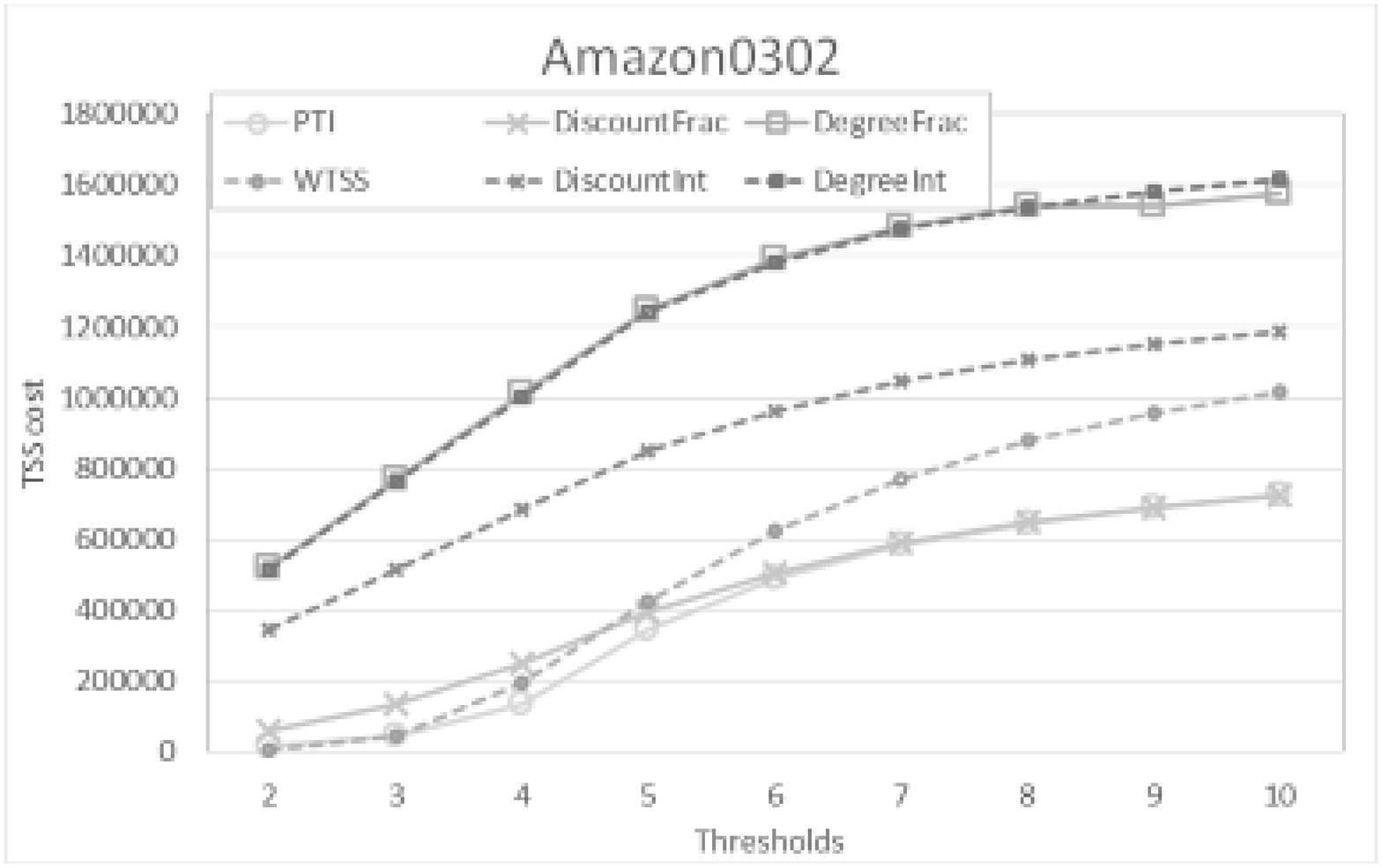}

\includegraphics[width=0.50\linewidth, height =6.3truecm]{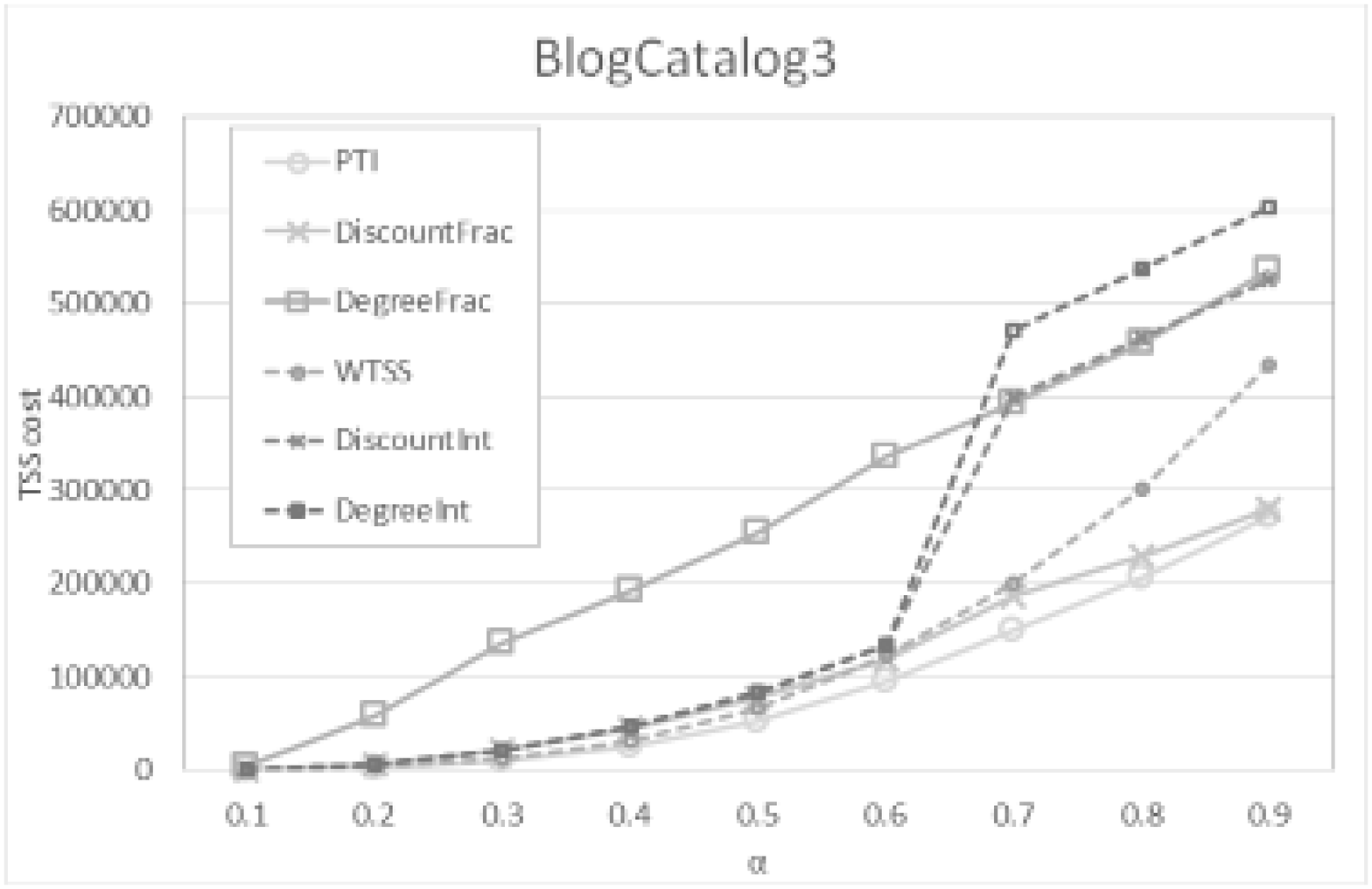}%
\includegraphics[width=0.50\linewidth, height =6.3truecm]{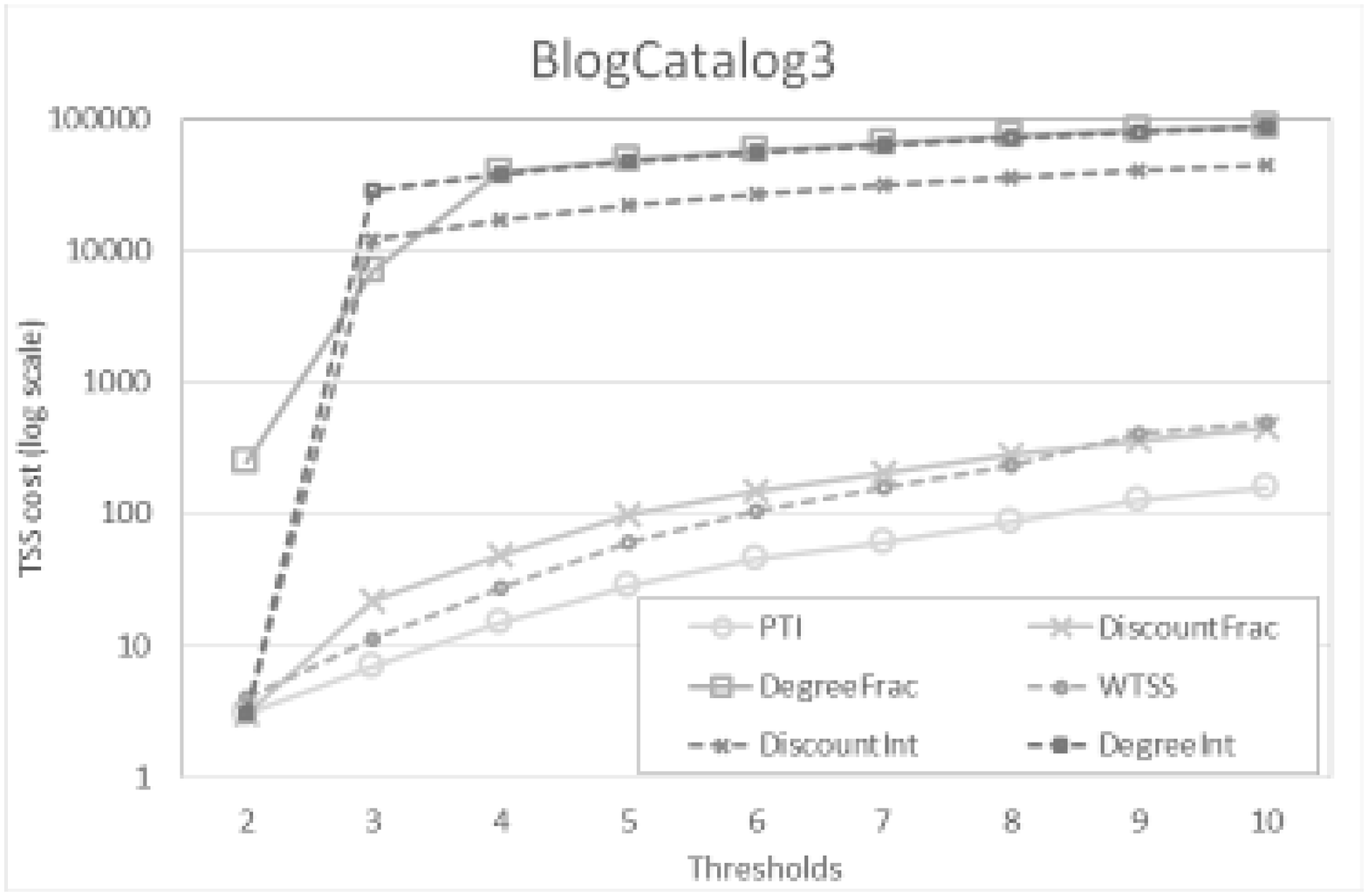}

\includegraphics[width=0.50\linewidth, height =6.3truecm]{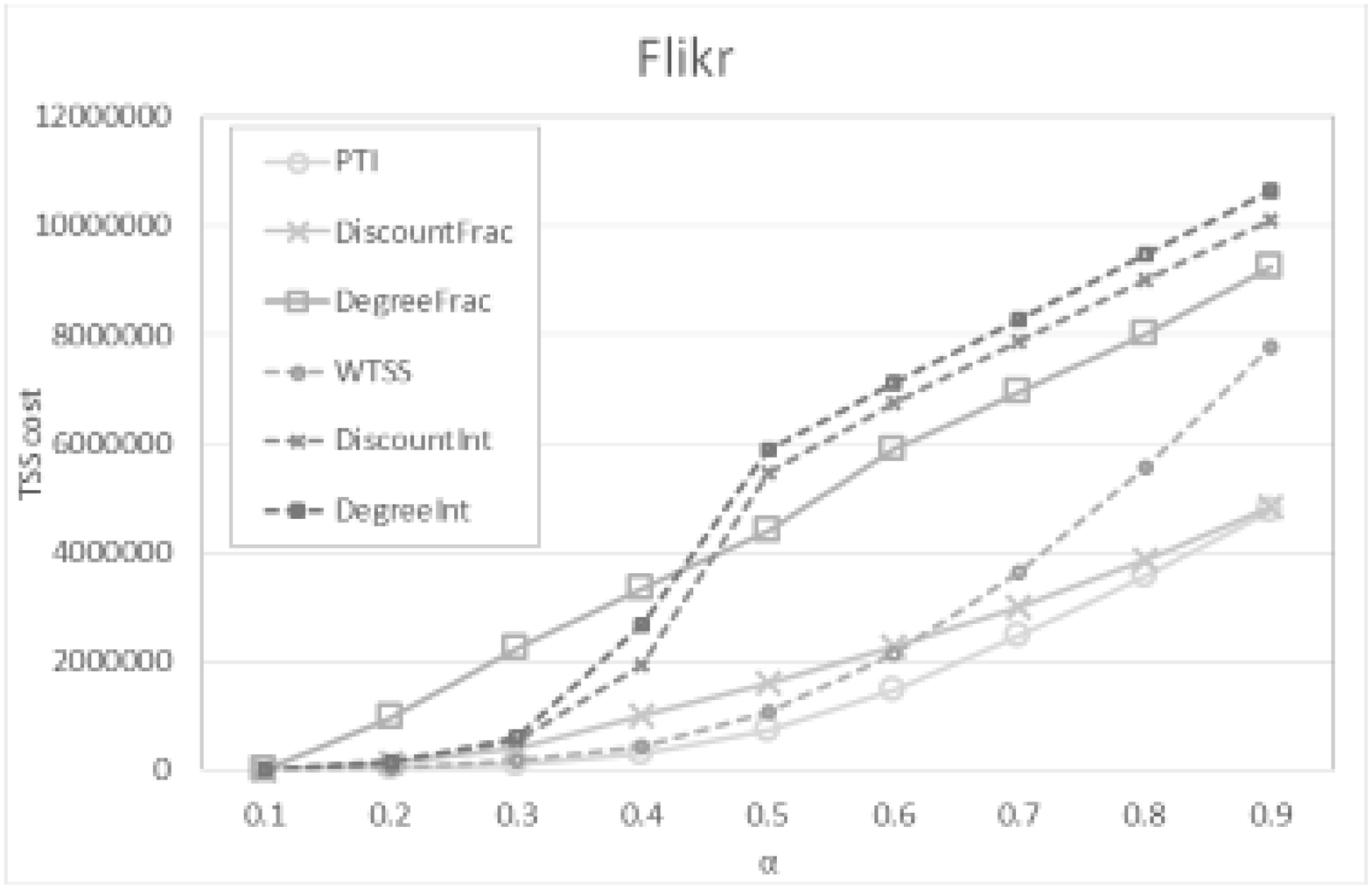}%
\includegraphics[width=0.50\linewidth, height =6.3truecm]{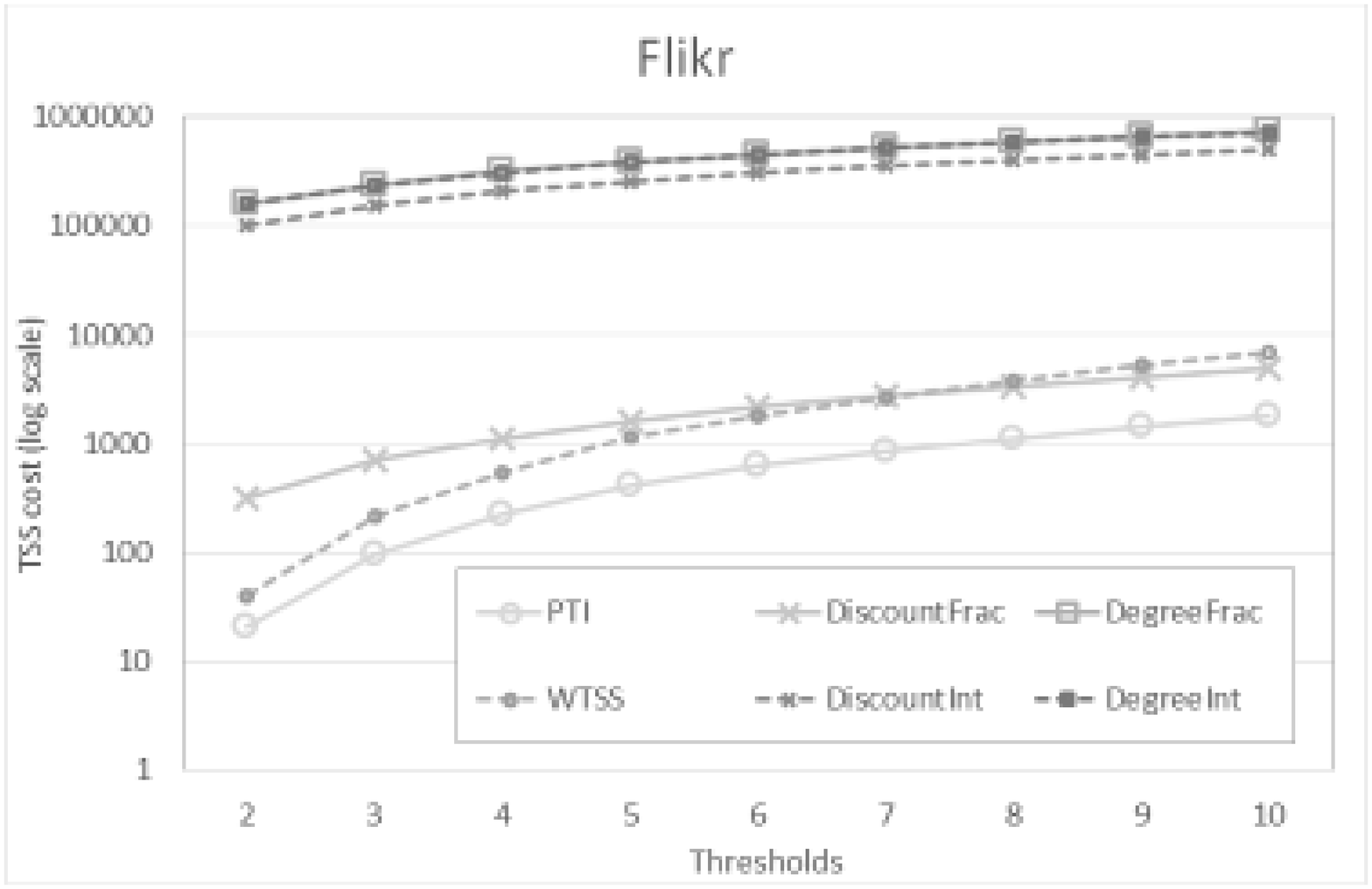}

\caption{Amazon0302, BlogCatalog3, and  Flikr 
results.} 
\label{Amazon}%
\end{center}
\end{figure}

\begin{figure}%
\begin{center}
\includegraphics[width=0.5\linewidth, height =6.3truecm]{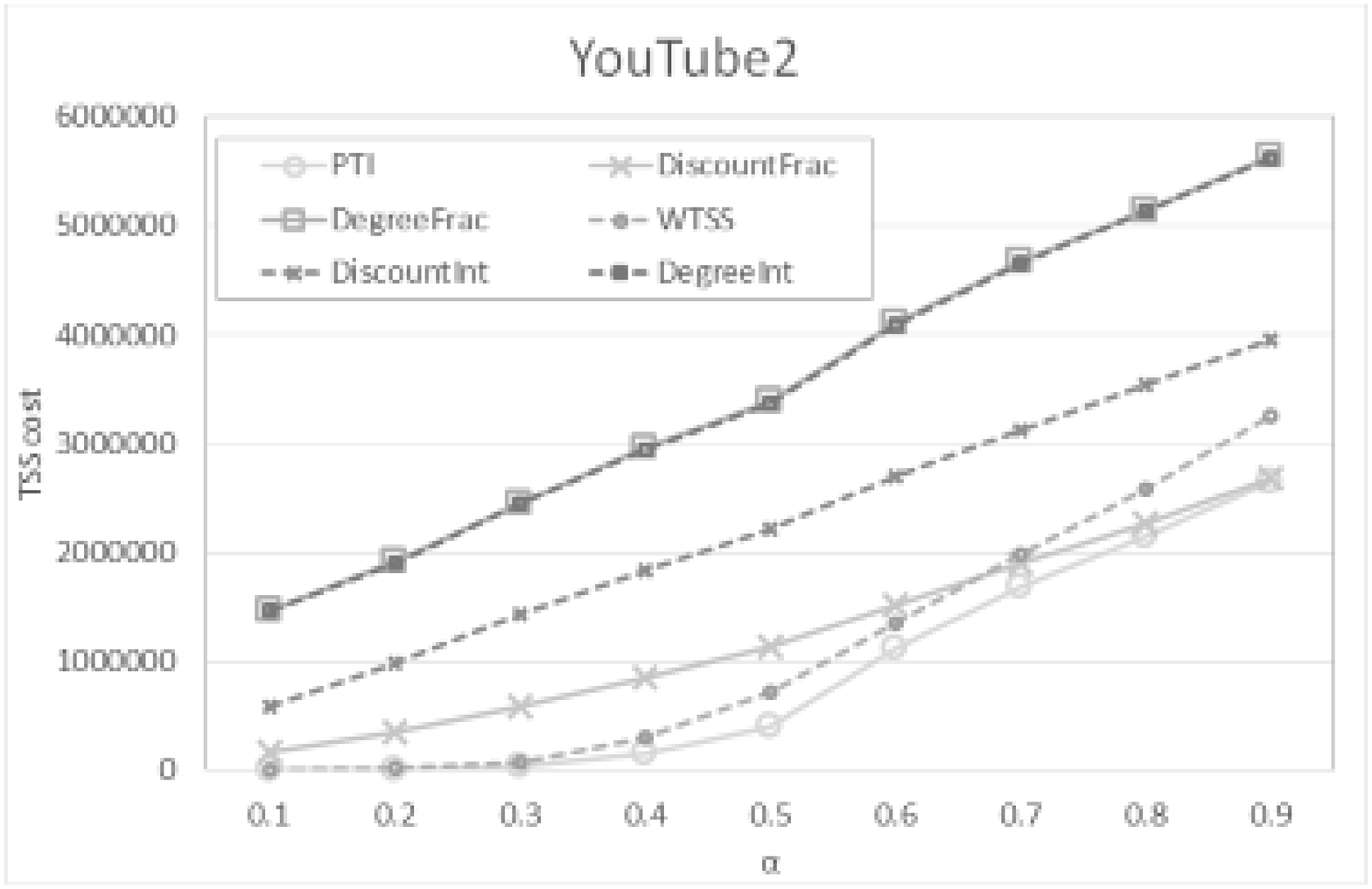}%
\includegraphics[width=0.5\linewidth, height =6.3truecm]{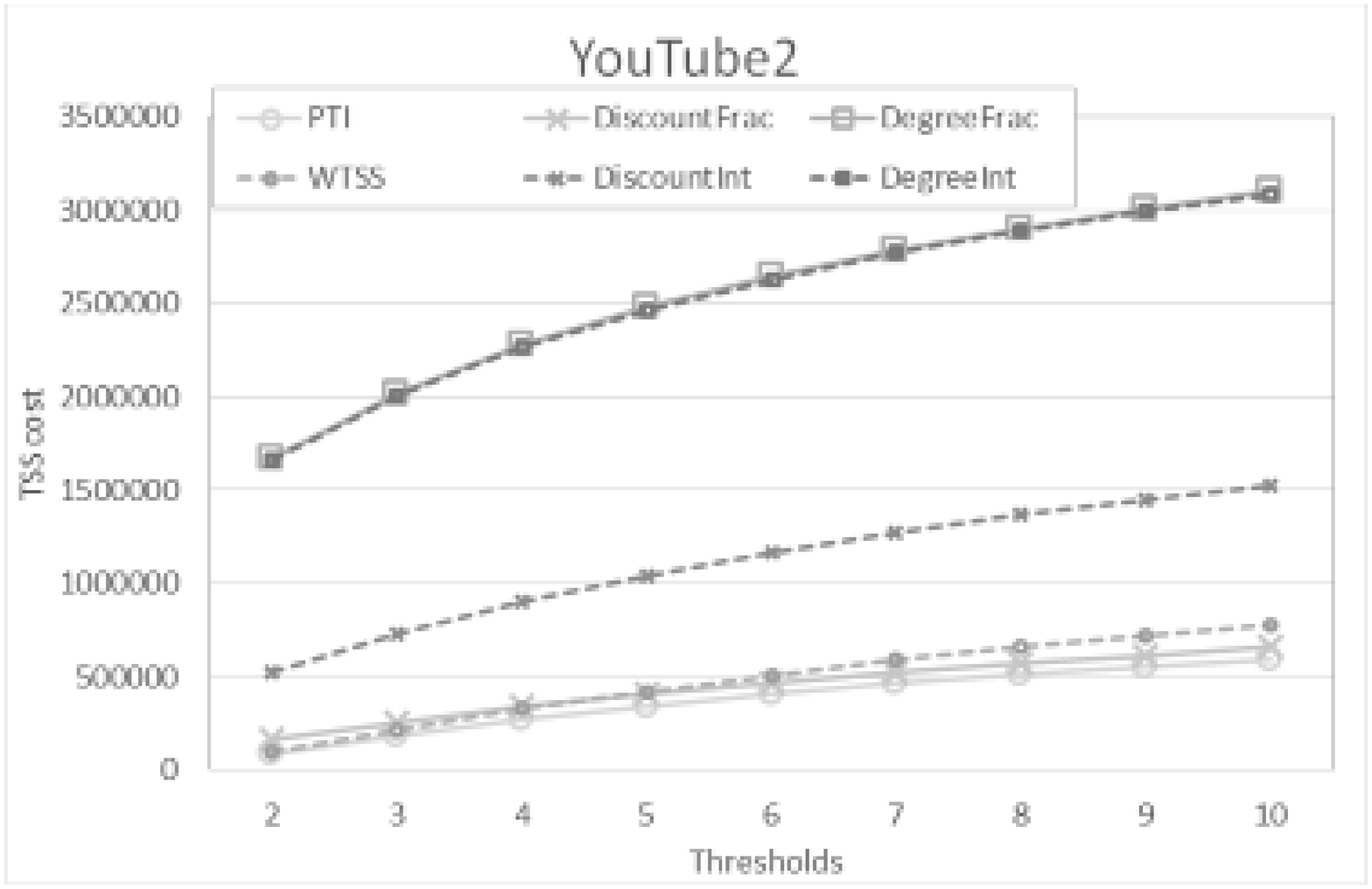}

\includegraphics[width=0.5\linewidth, height =6.3truecm]{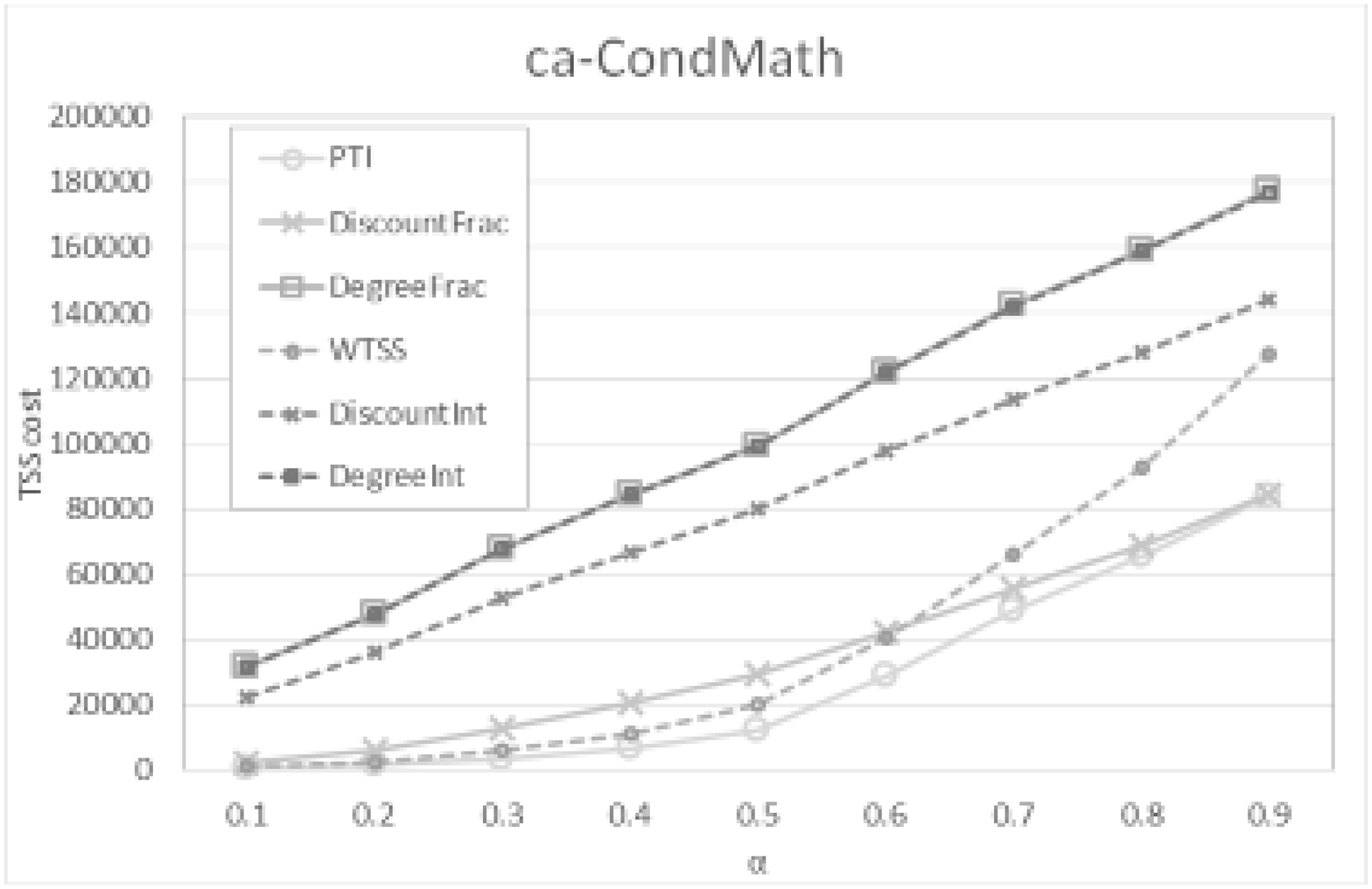}%
\includegraphics[width=0.5\linewidth, height =6.3truecm]{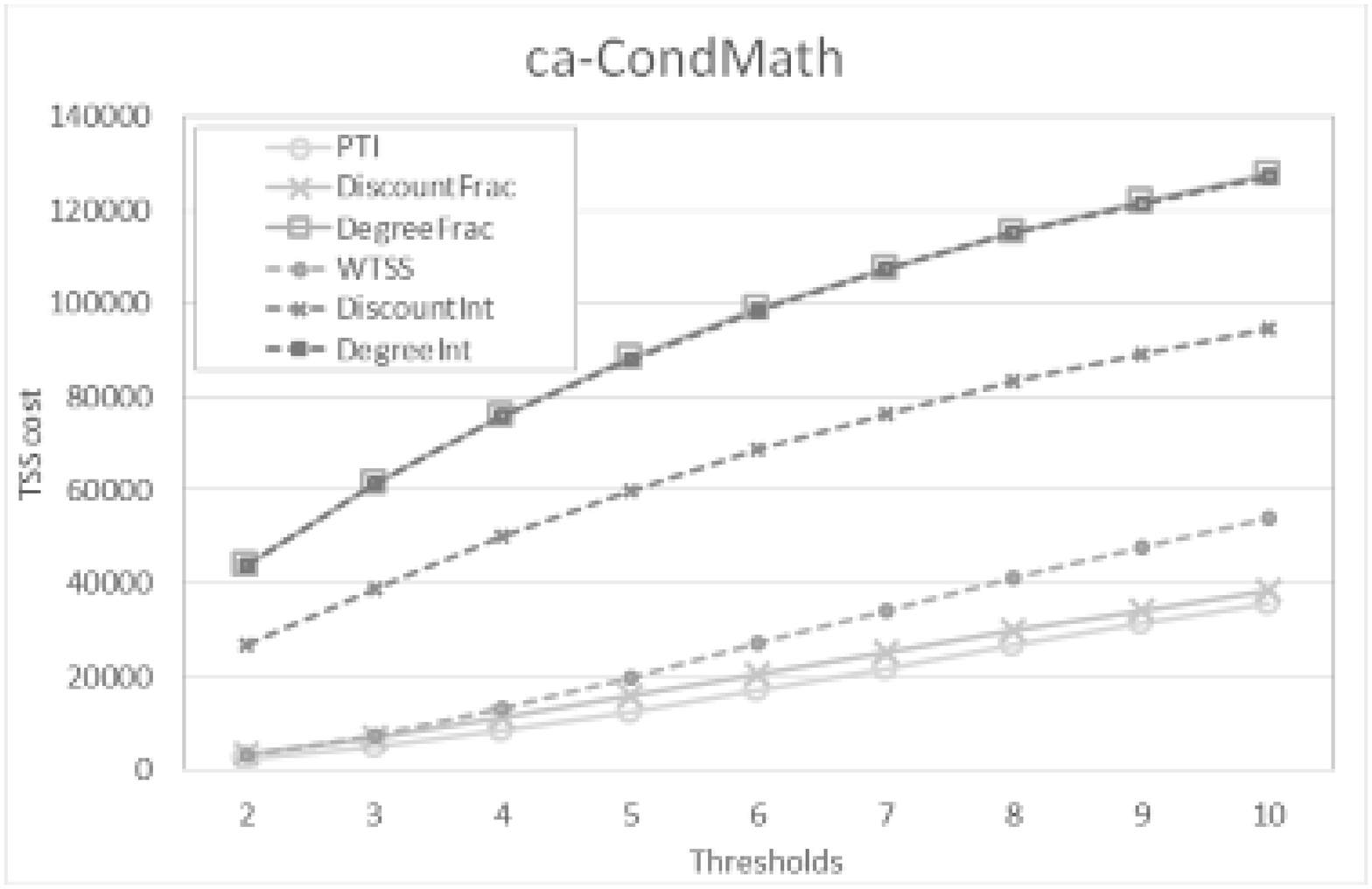}

\includegraphics[width=0.5\linewidth, height =6.3truecm]{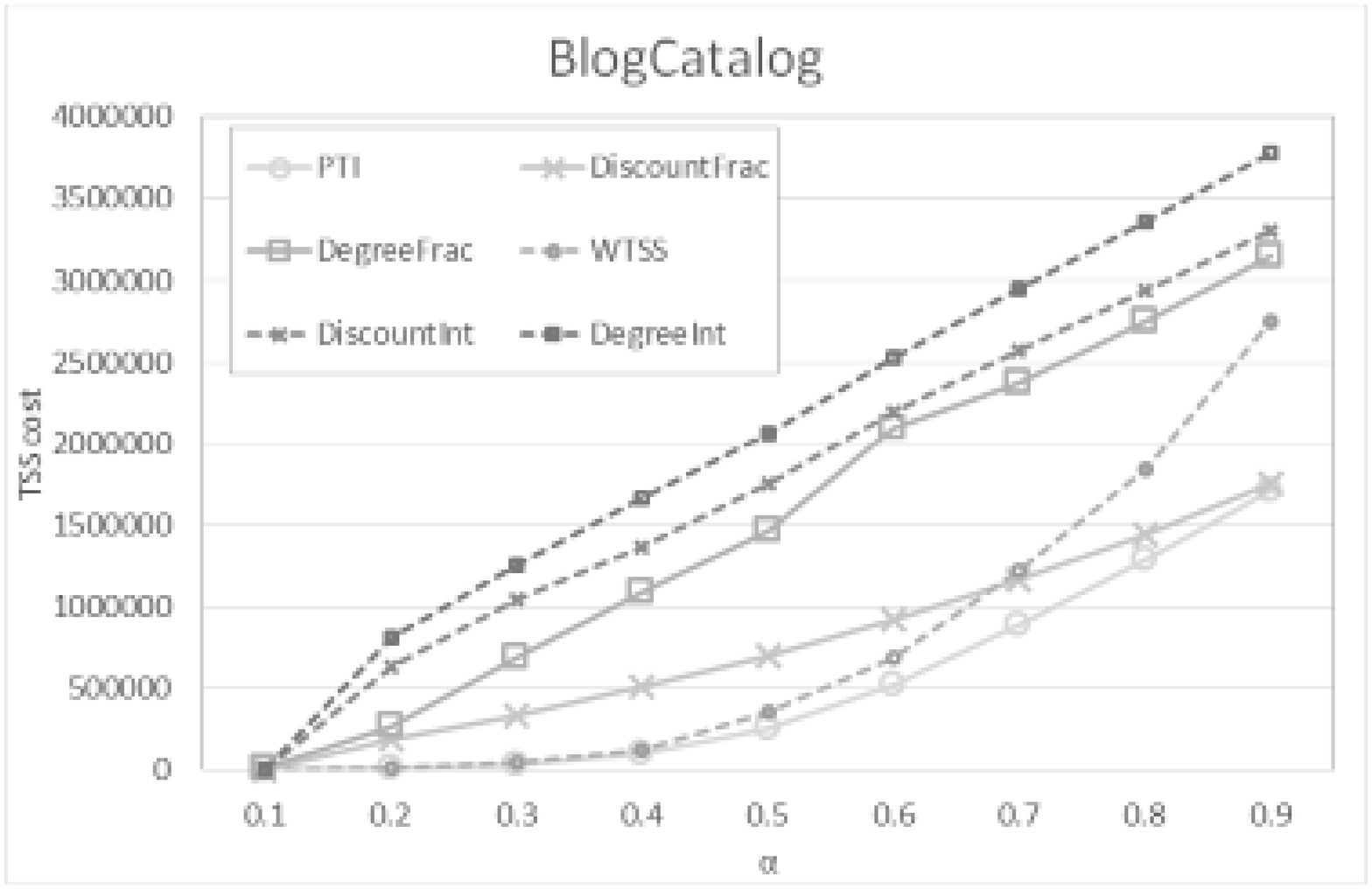}%
\includegraphics[width=0.5\linewidth, height =6.3truecm]{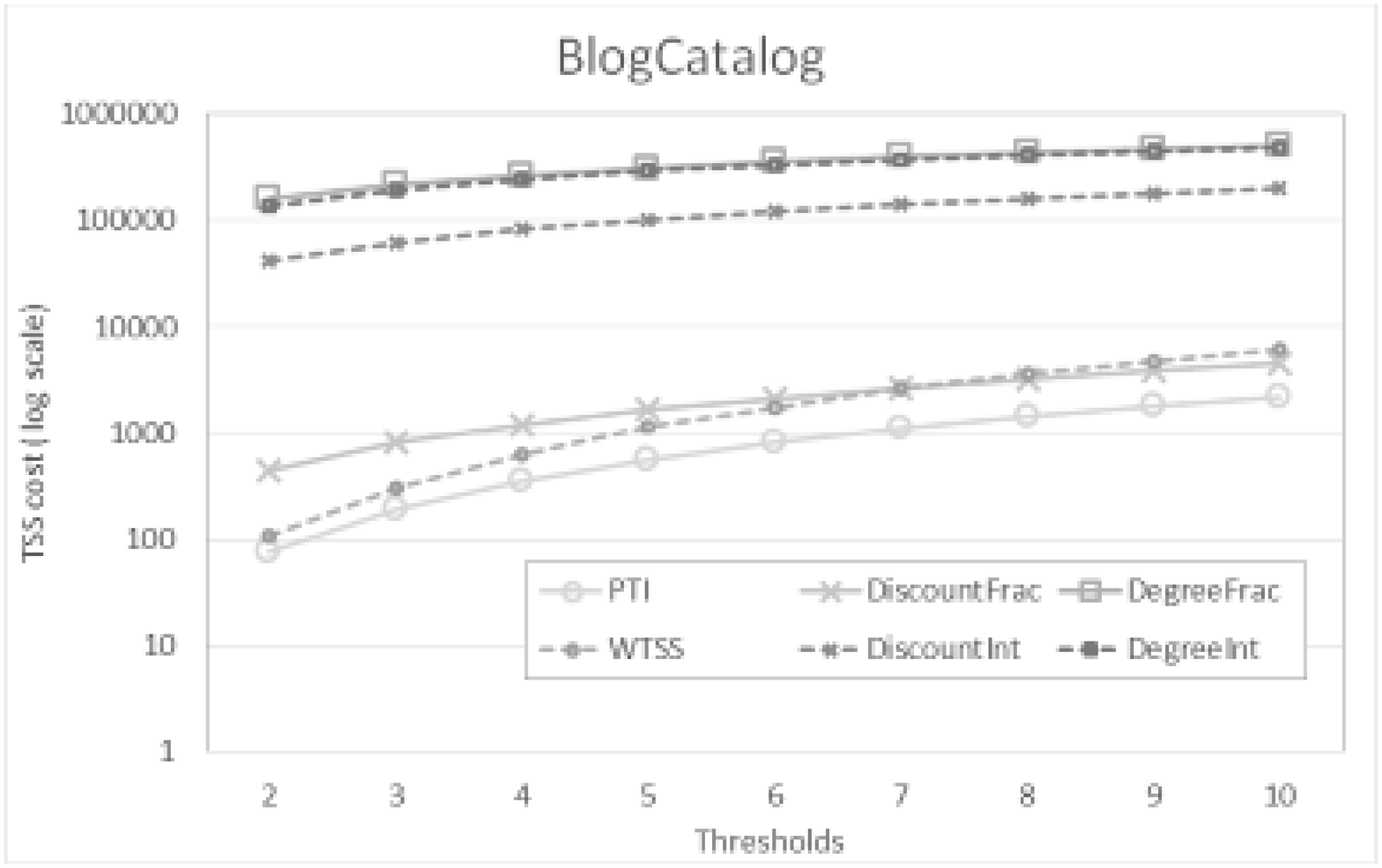}
\caption{ Youtube2, ca-CondMath,  and BlogCatalog    results.} 
\label{Amazon1}%
\end{center}
\end{figure}

\begin{figure}%
\begin{center}
\includegraphics[width=0.50\linewidth, height =6.3truecm]{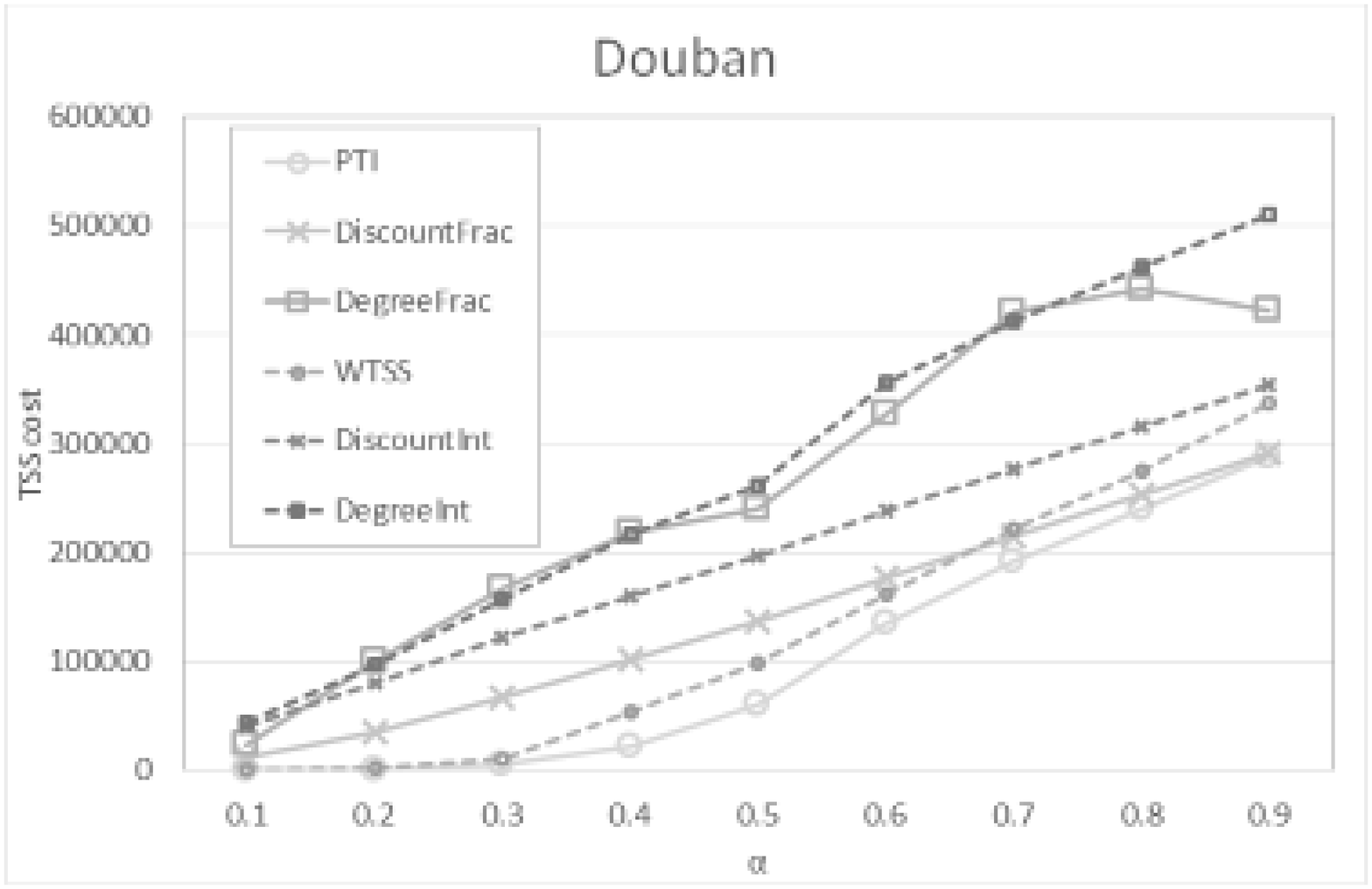}%
\includegraphics[width=0.50\linewidth, height =6.3truecm]{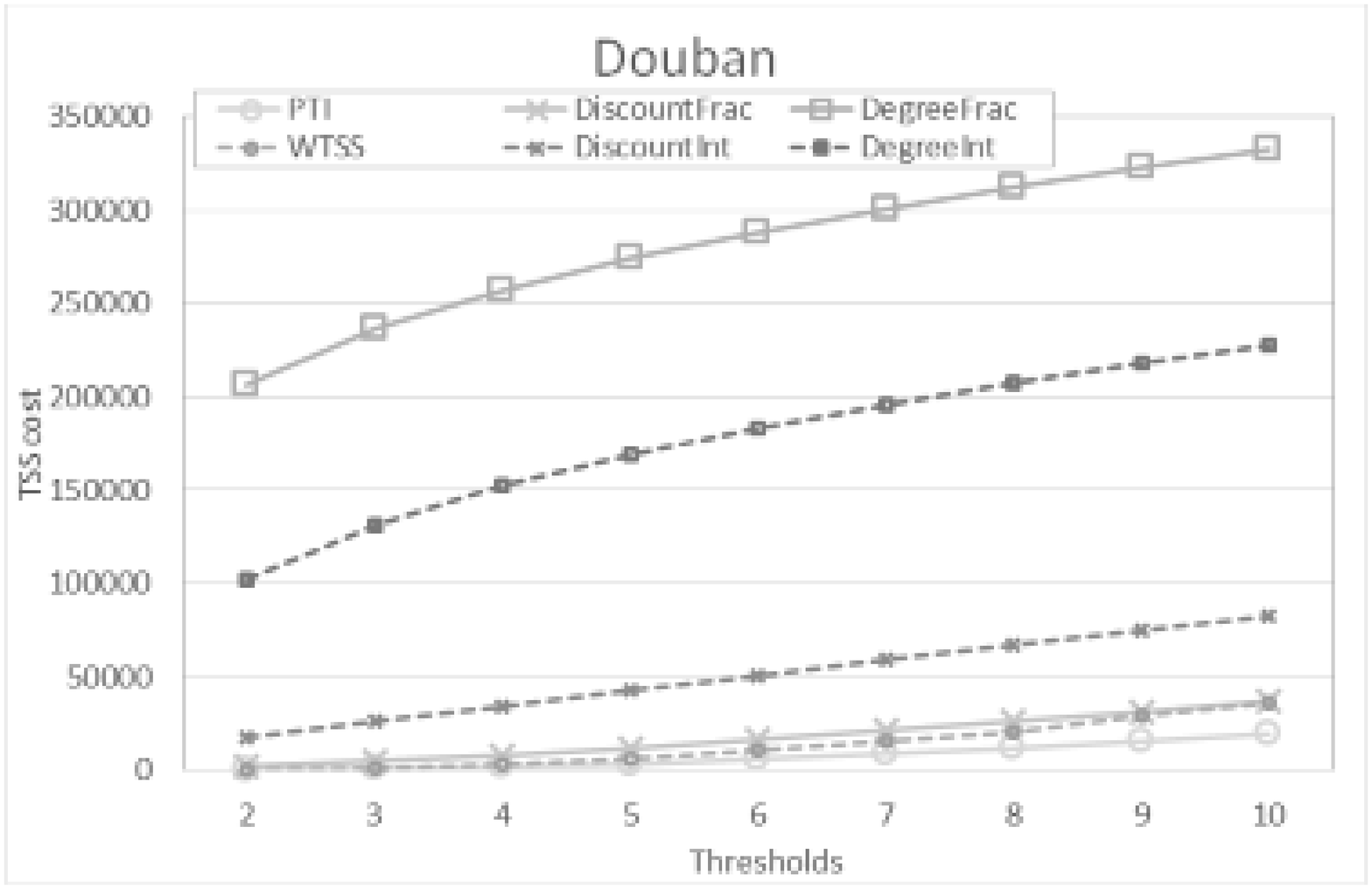}
\includegraphics[width=0.50\linewidth, height =6.3truecm]{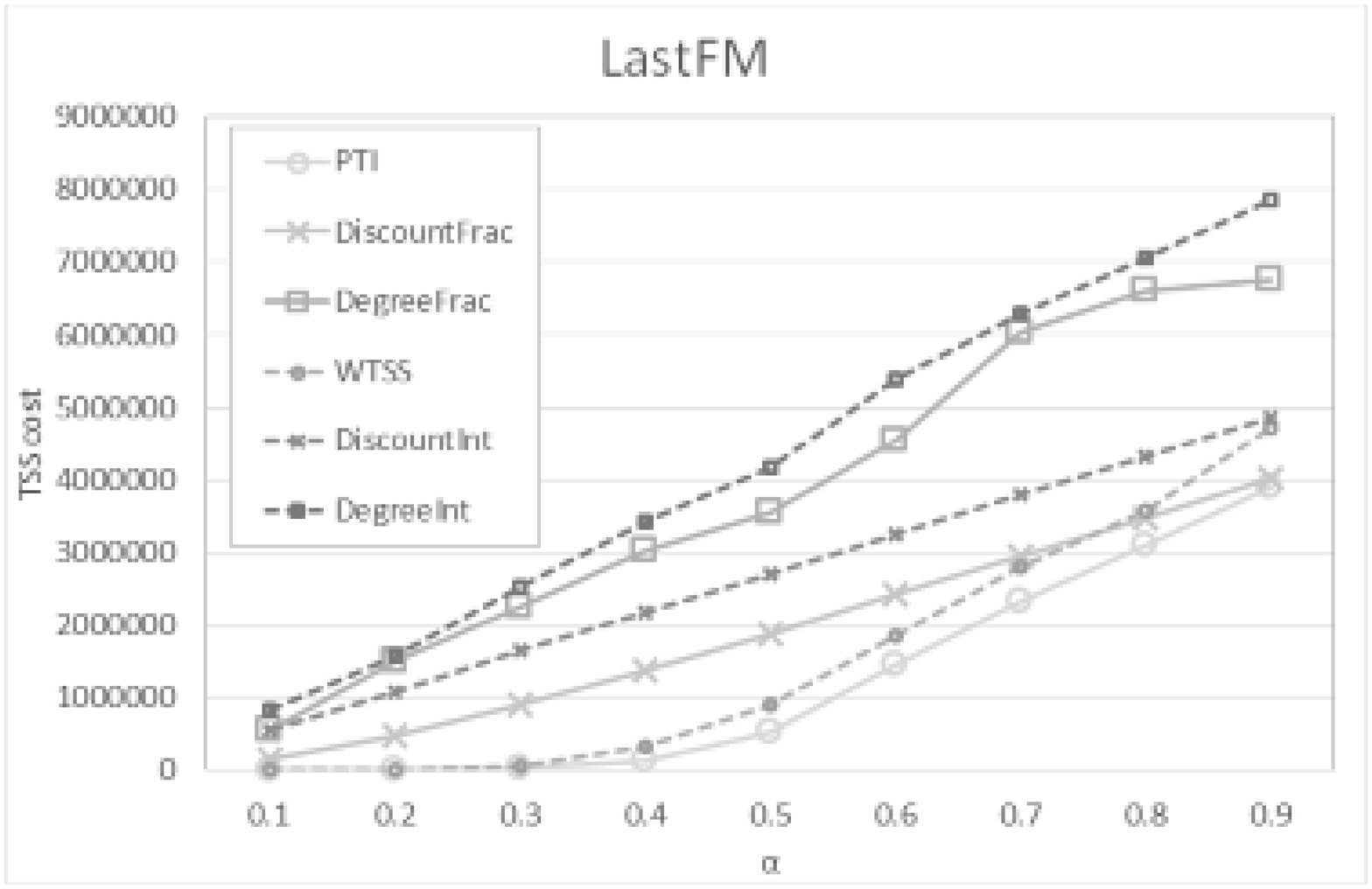}%
\includegraphics[width=0.50\linewidth, height =6.3truecm]{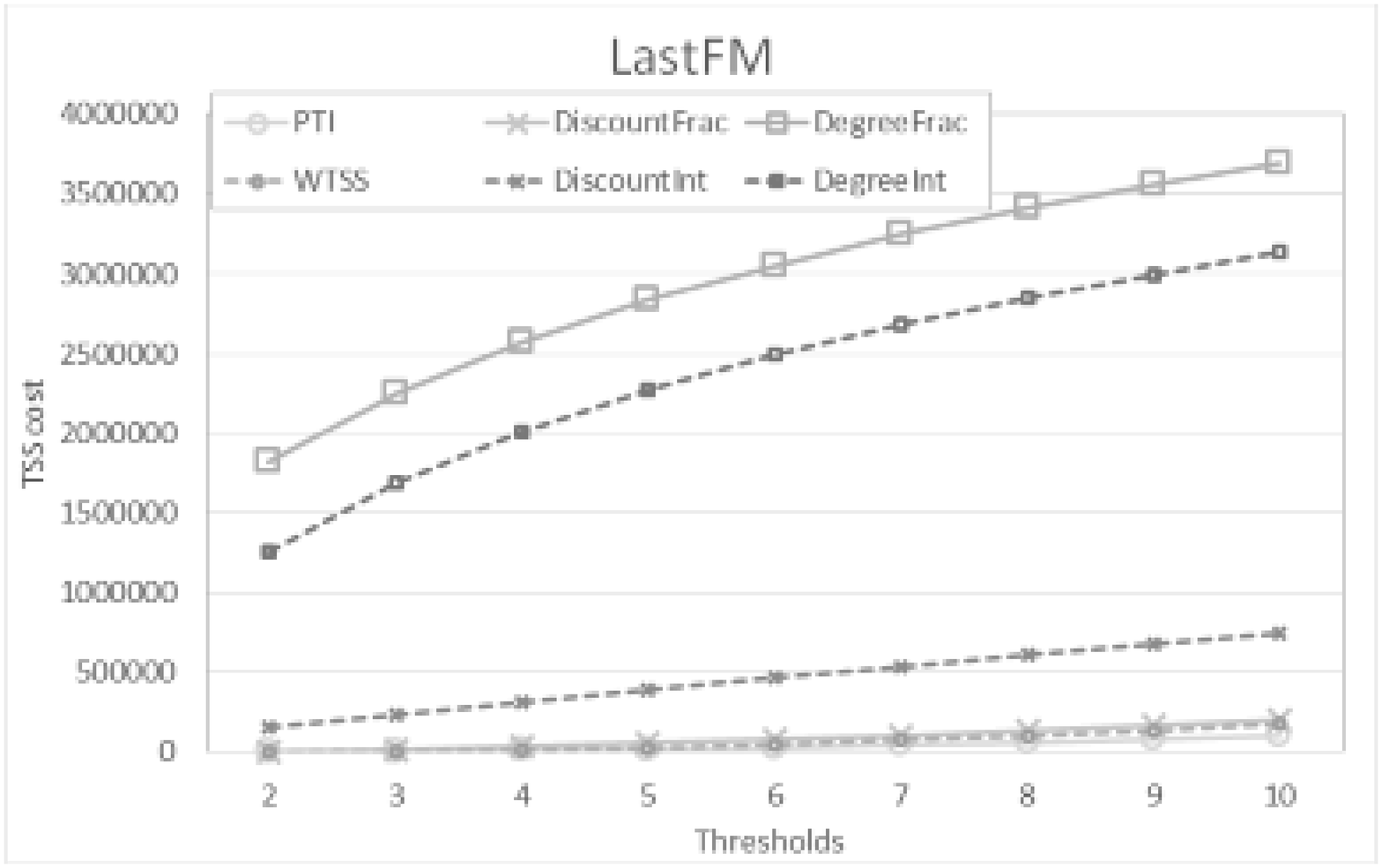}
\caption{
Douban and LastFM results.} 
\label{Amazon2}%
\end{center}
\end{figure}


\newpage

\nocite{41,Rei}

\baselineskip=0.55truecm

\bibliography{sirocco2015-TCS-arXiv}{}

\begin{thebibliography}{10}

\bibitem{ABW-10}
Eyal Ackerman, Oren Ben-Zwi, and Guy Wolfovitz.
\newblock Combinatorial model and bounds for target set selection.
\newblock {\em Theoretical Computer Science}, 411(44–46):4017 -- 4022, 2010.

\bibitem{Ba+}
Eytan Bakshy, Jake~M. Hofman, Winter~A. Mason, and Duncan~J. Watts.
\newblock Everyone's an influencer: Quantifying influence on twitter.
\newblock In {\em Proceedings of the Fourth ACM International Conference on Web
  Search and Data Mining}, WSDM '11, pages 65--74, New York, NY, USA, 2011.

\bibitem{BCNS}
Cristina Bazgan, Morgan Chopin, Andr\'e Nichterlein, and Florian Sikora.
\newblock Parameterized approximability of maximizing the spread of influence
  in networks.
\newblock {\em Journal of Discrete Algorithms}, 27:54 -- 65, 2014.

\bibitem{BHLM-11}
Oren Ben-Zwi, Danny Hermelin, Daniel Lokshtanov, and Ilan Newman.
\newblock Treewidth governs the complexity of target set selection.
\newblock {\em Discrete Optimization}, 8(1):87 -- 96, 2011.

\bibitem{Centeno12}
Carmen~C. Centeno, Mitre~C. Dourado, Lucia~Draque Penso, Dieter Rautenbach, and
  Jayme~L. Szwarcfiter.
\newblock Irreversible conversion of graphs.
\newblock {\em Theoretical Computer Science}, 412(29):3693 -- 3700, 2011.

\bibitem{Chang}
Ching-Lueh Chang.
\newblock Triggering cascades on undirected connected graphs.
\newblock {\em Information Processing Letters}, 111(19):973 -- 978, 2011.

\bibitem{Chen-09}
Ning Chen.
\newblock On the approximability of influence in social networks.
\newblock {\em SIAM Journal on Discrete Mathematics}, 23(3):1400--1415, 2009.

\bibitem{CLC}
W.~Chen, C.~Castillo, and L.~Lakshmanan.
\newblock {\em Information and Influence Propagation in Social Networks}.
\newblock Morgan \& Claypool, 2013.

\bibitem{CWY09}
Wei Chen, Yajun Wang, and Siyu Yang.
\newblock Efficient influence maximization in social networks.
\newblock In {\em Proceedings of the 15th ACM SIGKDD International Conference
  on Knowledge Discovery and Data Mining}, KDD '09, pages 199--208, New York,
  NY, USA, 2009.

\bibitem{Chun}
Chun-Ying Chiang, Liang-Hao Huang, Bo-Jr Li, Jiaojiao Wu, and Hong-Gwa Yeh.
\newblock Some results on the target set selection problem.
\newblock {\em Journal of Combinatorial Optimization}, 25(4):702--715, 2013.

\bibitem{Chun2}
Chun-Ying Chiang, Liang-Hao Huang, and Hong-Gwa Yeh.
\newblock Target set selection problem for honeycomb networks.
\newblock {\em SIAM Journal on Discrete Mathematics}, 27(1):310--328, 2013.

\bibitem{Chopin-12}
Morgan Chopin, Andr\'e Nichterlein, Rolf Niedermeier, and Mathias Weller.
\newblock Constant thresholds can make target set selection tractable.
\newblock {\em Theory of Computing Systems}, 55(1):61--83, 2014.

\bibitem{41}
Nicholas~A. Christakis and James~H. Fowler.
\newblock {The Collective Dynamics of Smoking in a Large Social Network}.
\newblock {\em N Engl J Med}, 358(21):2249--2258, May 2008.

\bibitem{CF}
Nicholas~A. Christakis and James~H. Fowler.
\newblock {\em {Connected: The Surprising Power of Our Social Networks and How
  They Shape Our Lives -- How Your Friends' Friends' Friends Affect Everything
  You Feel, Think, and Do}}.
\newblock Back Bay Books, reprint edition, January 2011.

\bibitem{Cic+}
Ferdinando Cicalese, Gennaro Cordasco, Luisa Gargano, Martin Milani\v{c},
  Joseph Peters, and Ugo Vaccaro.
\newblock Spread of influence in weighted networks under time and budget
  constraints.
\newblock {\em Theoretical Computer Science}, 586:40 -- 58, 2015.

\bibitem{Cic14}
Ferdinando Cicalese, Gennaro Cordasco, Luisa Gargano, Martin Milani\v{c}, and
  Ugo Vaccaro.
\newblock Latency-bounded target set selection in social networks.
\newblock {\em Theoretical Computer Science}, 535:1 -- 15, 2014.

\bibitem{C-OFKR}
Amin Coja-Oghlan, Uriel Feige, Michael Krivelevich, and Daniel Reichman.
\newblock Contagious sets in expanders.
\newblock In {\em Proceedings of the Twenty-Sixth Annual ACM-SIAM Symposium on
  Discrete Algorithms}, pages 1953--1987, 2015.

\bibitem{CGM+}
Gennaro Cordasco, Luisa Gargano, Marco Mecchia, Adele~Anna Rescigno, and Ugo
  Vaccaro.
\newblock {A Fast and Effective Heuristic for Discovering Small Target Sets in
  Social Networks}.
\newblock In {\em Proc. of 9th Annual International Conference on Combinatorial
  Optimization and Applications (COCOA 2015)}, volume LNCS 9486, pages
  193--208, 2015.

\bibitem{Dem14}
Erik~D. Demaine, Mohammad~Taghi Hajiaghayi, Hamid Mahini, David~L. Malec,
  S.~Raghavan, Anshul Sawant, and Morteza Zadimoghadam.
\newblock How to influence people with partial incentives.
\newblock In {\em Proceedings of the 23rd International Conference on World
  Wide Web}, WWW '14, pages 937--948, New York, NY, USA, 2014.

\bibitem{DR-01}
Pedro Domingos and Matt Richardson.
\newblock Mining the network value of customers.
\newblock In {\em Proceedings of the Seventh ACM SIGKDD International
  Conference on Knowledge Discovery and Data Mining}, KDD '01, pages 57--66,
  New York, NY, USA, 2001.

\bibitem{EK}
David Easley and Jon Kleinberg.
\newblock {\em Networks, Crowds, and Markets: Reasoning About a Highly
  Connected World}.
\newblock Cambridge University Press, New York, NY, USA, 2010.

\bibitem{FKRRS-2003}
Paola Flocchini, Rastislav Kr\'alovic, Peter Ruzicka, Alessandro Roncato, and
  Nicola Santoro.
\newblock On time versus size for monotone dynamic monopolies in regular
  topologies.
\newblock {\em Journal of Discrete Algorithms}, 1(2):129 -- 150, 2003.

\bibitem{Fr+}
D.~Freund, M.~Poloczek, and D.~Reichman.
\newblock Contagious sets in dense graphs.
\newblock In {\em Proceedings of 26th Int'l Workshop on Combinatorial
  Algorithms (IWOCA2015)}, 2015.

\bibitem{Ga+}
Luisa Gargano, Pavol Hell, Joseph~G. Peters, and Ugo Vaccaro.
\newblock Influence diffusion in social networks under time window constraints.
\newblock {\em Theor. Comput. Sci.}, 584(C):53--66, 2015.

\bibitem{Gr}
M.~Granovetter.
\newblock Threshold models of collective behavior.
\newblock {\em The American Journal of Sociology}, 83(6):1420--1443, 1978.

\bibitem{Gu+}
Alberto Guggiola and Guilhem Semerjian.
\newblock Minimal contagious sets in random regular graphs.
\newblock {\em Journal of Statistical Physics}, 158(2):300--358, 2015.

\bibitem{KKT-03}
David Kempe, Jon Kleinberg, and \'{E}va Tardos.
\newblock Maximizing the spread of influence through a social network.
\newblock In {\em Proceedings of the Ninth ACM SIGKDD International Conference
  on Knowledge Discovery and Data Mining}, KDD '03, pages 137--146, New York,
  NY, USA, 2003.

\bibitem{KKT-05}
David Kempe, Jon Kleinberg, and \'{E}va Tardos.
\newblock Influential nodes in a diffusion model for social networks.
\newblock In {\em Proceedings of the 32Nd International Conference on Automata,
  Languages and Programming}, ICALP'05, pages 1127--1138, Berlin, Heidelberg,
  2005.

\bibitem{snap}
J.~Leskovec and A.~Krevl.
\newblock {SNAP Datasets}: {Stanford} large network dataset collection.
\newblock http://snap.stanford.edu/data, 2015.

\bibitem{LAM}
Jure Leskovec, Lada~A. Adamic, and Bernardo~A. Huberman.
\newblock The dynamics of viral marketing.
\newblock {\em ACM Trans. Web}, 1(1), May 2007.

\bibitem{Li+}
Xianliang Liu, Zishen Yang, and Wei Wang.
\newblock Exact solutions for latency-bounded target set selection problem on
  some special families of graphs.
\newblock {\em Discrete Applied Mathematics}, 2015.

\bibitem{Mo+}
Flaviano Morone and Hernan~A. Makse.
\newblock {Influence maximization in complex networks through optimal
  percolation}.
\newblock {\em Nature}, 524(7563):65--68, June 2015.

\bibitem{N15}
M.~Newman.
\newblock Network data, {\tt http://www-personal.umich.edu/\textasciitilde
  mejn/netdata/}, 2015.

\bibitem{NNUW}
André Nichterlein, Rolf Niedermeier, Johannes Uhlmann, and Mathias Weller.
\newblock On tractable cases of target set selection.
\newblock {\em Social Network Analysis and Mining}, 3(2):233--256, 2013.

\bibitem{Re}
T.~V.~Thirumala Reddy and C.~Pandu Rangan.
\newblock Variants of spreading messages.
\newblock {\em J. Graph Algorithms Appl.}, 15(5):683--699, 2011.

\bibitem{Rei}
Daniel Reichman.
\newblock New bounds for contagious sets.
\newblock {\em Discrete Mathematics}, 312(10):1812 -- 1814, 2012.

\bibitem{W+}
Cheng Wang, Lili Deng, Gengui Zhou, and Meixian Jiang.
\newblock A global optimization algorithm for target set selection problems.
\newblock {\em Inf. Sci.}, 267:101--118, May 2014.

\bibitem{ZL09}
R.~Zafarani and H.~Liu.
\newblock Social computing data repository at {ASU}.
\newblock http://socialcomputing.asu.edu, 2009.

\bibitem{Za}
Manouchehr Zaker.
\newblock On dynamic monopolies of graphs with general thresholds.
\newblock {\em Discrete Mathematics}, 312(6):1136 -- 1143, 2012.

\end{thebibliography}
\bibliographystyle{plain}
\end{document}